\documentclass[12pt]{cmuthesis2}

\usepackage{graphicx,amsthm} % more modern
\usepackage{subfigure}
\usepackage{wrapfig}
\usepackage{sidecap}
\usepackage{authblk}
\usepackage{multicol}
\usepackage{algorithm}
\usepackage[noend]{algpseudocode}

\usepackage{bm}
\usepackage{amsmath}
\usepackage{amssymb}
\usepackage{amsfonts}
\usepackage{latexsym}
\usepackage{multirow,tikz,amsmath,comment}
\usepackage{xcolor}
\usepackage{ mathrsfs }

\usepackage{epstopdf}
\usepackage{centernot}
\usepackage{multirow,tikz,amsmath,comment}
\usetikzlibrary{arrows,fit,positioning,shapes,snakes}
\usepackage{enumitem}
\usetikzlibrary{positioning}

\usepackage{times}
\usepackage{fullpage}
\usepackage[numbers,sort]{natbib}
\usepackage[backref,pageanchor=true,plainpages=false, pdfpagelabels, bookmarks,bookmarksnumbered,
]{hyperref}
	
\newtheorem{theorem}{Theorem}

\newtheorem{lemma}{Lemma}
\newtheorem{Definition}{Definition}

\newcommand{\RNum}[1]{\uppercase\expandafter{\romannumeral #1\relax}}

\usepackage[letterpaper,twoside,vscale=.8,hscale=.75,nomarginpar,hmarginratio=1:1]{geometry}

\begin {document} 
\frontmatter

%initialize page style, so contents come out right (see bot) -mjz
\pagestyle{empty}

\title{ %% {\it \huge Thesis Proposal}\\
{\bf Diagnosis of Autism Spectrum Disorder by Causal Influence Strength Learned from Resting-State fMRI Data}}
%\author{Biwei Huang, Kun Zhang, Ruben Sanchez-Romero,  Joseph Ramsey, Madelyn Glymour, Clark Glymour}
\date{}
\Year{}
\trnumber{}

%\vspace{10cm}
 
\author{%
	\vspace{2cm}
	Biwei Huang$^\dagger$, Kun Zhang$^\dagger$, Ruben Sanchez-Romero$^\star$,  Joseph Ramsey$^\dagger$, Madelyn Glymour$^\dagger$, Clark Glymour$^\dagger$ \\
	$^\dagger$ Department of Philosophy, Carnegie Mellon University\\
    $^\star$Center for Molecular and Behavioral Neuroscience, Rutgers University  \\
    {\{biweih, kunz1, jdramsey, mglymour, cg09\}@andrew.cmu.edu, ruben.saro@gmail.com}\\
	% examples of more authors
%	 \And
%	 Kun Zhang \\
%	 Affiliation \\
%	 Address \\
%	 \texttt{email} \\
%	 \AND
%	 Ruben Sanchez-Romero \\
%	 Affiliation \\
%	 Address \\
%	 \texttt{email} \\
%	 \And
%	 Joseph Ramsey \\
%	 Affiliation \\
%	 Address \\
%	 \texttt{email} \\
%	 \And
%	 Madelyn Glymour \\
%	 Affiliation \\
%	 Address \\
%	 \texttt{email} \\
}
%\support{}
%\disclaimer{}

% copyright notice generated automatically from Year and author.
% permission added if \permission{} given.

%\keywords{ASD diagnosis; Causal discovery; Resting-state fMRI.}

\maketitle

\pagestyle{plain} % for toc, was empty

%% Obviously, it's probably a good idea to break the various sections of your thesis
%% into different files and input them into this file...

\begin{abstract}
Autism spectrum disorder (ASD) is one of the major developmental disorders affecting children; around 1 in 68 children in United States has been identified as some form of ASD. With the hard situations individuals with ASD may face, it is urgent to diagnose early and accurately and then provide proper service and treatment to them. Current clinical diagnoses mainly rely on descriptions and observations of behavior. However, behavior assessment has limitations in the aspect that it is subject to human judgment highly depending on doctors' experiences. More importantly, behavior assessment is unable to find out the underlying biological bases behind the observed behavioral symptoms, while knowledge of such biological bases may help to find a proper treatment.

Recently, with the hypothesis that ASD is associated with atypical brain connectivities, several studies investigate the differences in brain connectivities between ASD and typical controls from functional Magnetic Resonance Imaging (fMRI). A substantial body of researches use Pearson's correlation coefficients, mutual information, or partial correlation to estimate statistical dependencies between brain areas from neuroimaging data. However, correlation or partial correlation does not directly reveal causal influences—the information flow—between brain regions. Compared to correlation, causality pinpoints the key connectivity characteristics and removes some redundant features for diagnosis. It also gives the direction of information flow.

In this paper, we propose a two-step method for large-scale and cyclic causal discovery from fMRI data. It can identify brain causal structures without doing interventional experiments. Particularly, the proposed approach to causal discovery relies on constrained functional causal models,
which represents the effect as a linear function of the direct causes and independent noise terms. It is able to recover the whole causal structure between brain regions and allow feedbacks. The learned causal structure, as well as the causal influence strength, provides us the path and effectiveness of information flow, i.e., whether region A directly influences region B, and if there is a direct influence, how strong it is. With the recovered causal influence strength as candidate features, we then perform ASD diagnosis by further doing feature selection and classification. We apply our methods to three datasets from Autism Brain Imaging Data Exchange (ABIDE). 

From experimental results, it shows that with causal connectivities, the diagnostic accuracy largely improves, compared with that using correlation and partial correlation as measures. A closer examination shows that information flow starting from the superior front gyrus to other areas are largely reduced; the target areas are mostly concentrated in default mode network and the posterior part. Moreover, all enhanced information flows are from posterior to anterior or in local areas, while local reductions also exist. Overall, it shows that long-range influences have a larger proportion of reductions than local ones, while local influences have a larger proportion of increases than long-range ones. By examining the graph properties of brain causal structure, the group of ASD shows reduced small-worldness.
\end{abstract}

\tableofcontents
\listoffigures
\listoftables

\mainmatter

\begin{Huge}
	 \textbf{Abbreviations}
\end{Huge}
\begin{table}[htp!]
	\centering
	%\caption{My caption}
	\begin{tabular}{llll}
		ASD    &  &  & Autism Spectrum Disorder                            \\
		TC     &  &  & Typical Control                                     \\
		DSM    &  &  & Diagnostic and Statistical Manual                   \\
		ADI-R  &  &  & Autism Diagnostic Interview-Revised                 \\
		ADOS   &  &  & Autism Diagnostic Observation Schedule              \\
		fMRI   &  &  & functional Magnetic Resonance Imaging               \\
		MRI    &  &  & Magnetic Resonance Imaging                          \\
		ABIDE  &  &  & Autism Brain Imaging Data Exchange                  \\
		ROI    &  &  & Regions of Interest                                 \\
		LiNGAM &  &  & Linear Non-Gaussian Acyclic model                   \\
		FCI    &  &  & Fast Causal Inference                               \\
		CCD    &  &  & Cyclic Causal Discovery                             \\
		DAG  &  &  &  Directed Acyclic Graph                           \\
		PAG    &  &  & Partial Ancestral Graph                             \\
		GES    &  &  & Greedy Equivalence Search                           \\
		FCM    &  &  & Functional Causal Model                             \\
		ICA    &  &  & Independent Component Analysis                      \\
		SCAD   &  &  & Smoothly Clipped Absolute Deviation                 \\
		GSCAD  &  &  & Generalized Smoothly Clipped Absolute Deviation     \\
		R-fMRI &  &  & Resting-state functional Magnetic Resonance Imaging \\
		CCS    &  &  & Connectome Computation System                       \\
		AAL    &  &  & Automated Anatomical Labeling                       \\
		SVM    &  &  & Support Vector Machine                             
	\end{tabular}
\end{table}

\chapter{Introduction}

\section{What is Autism Spectrum Disorder}
Autism spectrum disorder (ASD) is a family of developmental disorders, including autistic disorder, asperger syndrome, pervasive developmental disorder--not otherwise specified, and childhood disintegrative disorder. According to the Diagnostic and Statistical Manual (DSM-5) \cite{DSM5}, it is defined as the presence of deficits or unusual behaviors within three domains: reciprocal social interactions, communication deficits, and restricted, repetitive interests and behaviors. The onset of these unusual behaviors is prior to 3 years old. ASD is one of the major problems affecting children. As shown by the Centers for Disease Control and Prevention, around 1 in 68 children in United States has been identified as some form of ASD \cite{CDC}. 

It has been shown that many individuals with ASD exhibit impairments in learning, development, as well as daily life skills. (1) It may affect development. Children with ASD develop at a different rate and do not necessarily develop skills in the same order as typically developing children. For example, a child with ASD might be unable to combine words together into short phrases not until three years old. (2) It may affect attention and interaction. Children with ASD may not tune into other people in the same way as typically developing ones. For example, a child with ASD might not respond to his name, make eye contact, smile at caregivers, or wave goodbye without being told to. A child with ASD also might not know how to get others' attention or communicate with others. (3) It may affect understanding. Children with ASD usually find it hard to see things from other people's perspectives. They might have trouble understanding that other people can have different desires and beliefs from them. They might also find it hard to understand and predict other people's behavior, and to understand how their behavior affects others. (4) It may affect control and regulation. Children with ASD may struggle with focus, attention, transitions, organization, memory, time management, emotional control, and frustration. (5) It may affect to see the `big picture'. Children with ASD have difficulty seeing the `big picture'. They can get lost in the details, rather than pulling together different sources of information and seeing the situation as a whole.

\section{Autism Spectrum Disorder Diagnosis}
With the hard situations individuals with ASD may face, it is urgent to diagnose early and accurately, and then provide proper service and treatment to them.
Current clinical diagnoses of ASD mainly rely on descriptions and observations of behavior. The most-widely used diagnostic criteria are the Autism Diagnostic Interview-Revised (ADI-R) \cite{ADIR} and the Autism Diagnostic Observation Schedule (ADOS) \cite{ADOS}. However, behavior assessment has limitations in several aspects. First, the decision  from human judgment, whether the subject has ASD or not, is subjective; it highly depends on doctors' experiences. Second, it may need a relatively long period's observation and evaluation to make a decision. This process may include a number of different professionals from different fields. Diagnosis of ASD in adults is even harder. In adults, some ASD symptoms overlap with those of other mental health disorders, such as schizophrenia or attention deficit hyperactivity disorder. Third, the key point is that behavior assessment is unable to find out the underlying biological bases behind the observed behavioral symptoms. Knowledge of such biological bases may help to find a proper treatment. For example, if we can identify genetic causes or pathophysiology of ASD, it may motivate the development of corresponding medical treatment.

%(introduce fMRI)
In the past decades, the development of neuroimaging techniques provides us a noninvasive way to measure whole brain activities and then to infer neuropathology biomarkers of ASD. Particularly, the functional magnetic resonance imaging (fMRI) works by detecting the changes in blood oxygenation and flow that occur in response to neural activities; when a brain area is more active, it consumes more oxygen, and to meet this increased demand, blood flow increases to the active area. The fMRI has several significant advantages: it is noninvasive and does not involve radiation, making it safe for the subject; it has excellent spatial and good temporal resolution; it is easy for the experimenter to use. The attractions of fMRI have made it a popular tool for imaging brain functions.  According to whether there is a specific task or stimulus, there are two types of fMRI: tasked-based fMRI and resting-state fMRI. Compared to task-based fMRI, resting-state fMRI provides a promising way to localize neuropathology diagnostic biomarkers with the following advantages \cite{Resting-task}. 1). It can be used to study various systems (e.g., language and motor systems), while task fMRI needs different experimental designs to study different systems. 2). It is relatively fast to acquire and requires minimum cooperation from the patients.

There have been a number of research studies on ASD diagnosis from neuroimaging data. Different types of features have been considered and extracted, e.g., anatomical structures, neural activities, and neural connectivities. Preliminary morphology studies have reported reduced volume in thalamus \cite{ASD_thalamus} and amygdala \cite{ASD_amygdala1, ASD_amygdala2, ASD_amygdala3} in individuals with autism compared with control group. The local cortical thickness measure has been shown to be correlated with symptom severity in ASD and has been used as a predictor for it \cite{ASD_thick}.
Preliminary functional studies focusing on regional brain activation in autism have reported abnormal activities in a diverse set of brain regions, e.g., reduced activation in the amygdala when processing facial expressions \cite{ASD_amygdala4, ASD_amygdala5}, reduced activation within frontal, parietal, and occipital regions during a spatial attention task \cite{ASD_attention1}.

Recently, with the hypothesis that ASD is associated with atypical brain connectivities, several studies consider the differences in brain connectivities between ASD and typical controls. Hypoconnectivity between amygdala and cortical regions \cite{ASD_amygdala6}, and between amygdala and insula \cite{ASD_insula1} have been found in adolescents and adults with autism compared with healthy controls.  Altered functional connectivities from the thalamus to the cortex could lead to impaired information transmission from the sensory periphery to the cortex in individuals with ASD \cite{ASD_thalamus2}. Research examining developmental changes in large-scale network functional connectivity reported reduced connectivity between amygdala/subcortical network and default mode network in adolescents with autism, as well as increased connectivity within amygdala/subcortical network in children with autism \cite{ASD_amygdala7}. It has also been suggested that the brains of individuals with autism exhibit reduced long-range connectivity \cite{Marcel_Just2}.

However, some of the findings are not consistent across different research groups. According to a summary in \cite{ASD_network4}, the posterior cingulate cortex has been reported in eight studies to have reduced, and in three reports to have increased, long-range connectivity to other ROIs. The precuneus, anterior cingulate cortex, superior temporal gyrus, posterior superior temporal sulcus, anterior insula, and parietal lobule each has four to five reports of reduced, with two reports of increased, long-range connectivities. 

The diversity may come from several sources. (1) Small sample size: In most previous studies, the number of subjects is around 15 to 20 for both ASD and control groups, and sometimes it is even smaller. The small sample size may lead to large variance of identified biomarkers and thus less reliability. (2) Combined subtypes: Since ASD is a family of developmental disorders, it has several subtypes. Different studies may include subjects from different subtypes, and most of them do not distinguish between them.  Furthermore, behavior assessment of different subtypes might not be accurate, and even the label information from behavior diagnosis may have noise. (3) Different machines and experimental protocols are used to collect the data. Moreover, there are large differences in data analysis approaches in different studies. Even different preprocessing procedure may lead to large variability.

There is an urgent need to deal with the above problems, if we are to discover clinically useful information. To solve the small-sample-size problem, datasets with larger sample sizes are needed. Large samples may allow the extraction of core abnormalities from the noise introduced by the heterogeneity of the disorder. Such abnormalities could serve as biomarkers and could provide insight into the causes of the disorder and potential interventions.
The Autism Brain Imaging Data Exchange (ABIDE) \cite{ABIDE1} brings together neuroimaging data from multiple sites, acquired on multiple types of scanners, and with differing protocols. It provides previously collected datasets composed of both MRI data and phenotype information from 16 different international sites for over 1100 individuals, approximately half of whom are typically developing and half are diagnosed as ASD. This sample size, which is more than an order of magnitude larger than that used in most single-site studies, provides the power needed to identify neuroanatomical abnormalities related to ASD, although the multi-site, multi-protocol aspect of the data introduces additional heterogeneity. 

Regarding reliable data analysis approaches, in particular, the approach to identify brain connectivities from neuroimaging data, a substantial body of researches uses Pearson correlation coefficients, mutual information, or partial correlation.
In \cite{ASD_amygdala5, ASD_amygdala6, ASD_amygdala7}, functional connectivities ware computed for each participant and each pair of regions of interest (ROIs) as a correlation between the average time courses of all the voxels of ROIs. To reliably estimate the true covariance matrix with limited time points for a given individual, the Ledoit-Wolf shrinkage estimator, a parameter-free regularized covariance estimator, is used to estimate relationships between ROI time series \cite{ASD_network8}. There are also studies which use partial correlation as a measurement of brain connectivities \cite{Autism_partial1, Autism_partial2}.  However, correlation or partial correlation does not directly reveal causal influences—the information flow—between brain regions. Compared to correlation, causality pinpoints the key connectivity characteristics and removes some redundant features for diagnosis. It also gives the direction of information flow. For example, suppose that the information flow between three brain regions is $A \leftarrow C \rightarrow B$, if we use correlation as a measure, we will get a spurious edge between $A$ and $B$, and we cannot derive the direction between these three regions. 

With benefits from using causal connectivity compared to association measure, then one might ask how we can infer causal connectivities. Traditionally, to infer causal connectivity, interventions or randomized experiments are often used. However, they may be expensive or even impossible to do. For example, one wants to know which neurons lead to rat's whisker twitch. Suppose a rat's motor cortex contains 1 million neuron, and there are 100 neurons directly lead to whisker twitch. How can we find out these 100 neurons? If we do randomized experiments, then we need to do numerous expensive experiments, which is impossible in practice. 

Recent progresses in causal discovery make it possible to discover causal relationships from observational data, without doing interventions or randomized experiments. Roughly speaking, there are two types of well-developed methods in causal discovery. One is the constraint-based method. Under appropriate assumptions, it recovers a Markov equivalence class of the underlying causal structure based on conditional independence relationships among the variables \cite{SGS}. The resulting equivalence class may contain multiple directed acyclic graphs, which entail the same conditional independence relationships. The required assumptions include the causal Markov condition and the faithfulness assumption, which entail a correspondence between separation properties in the underlying causal structure and statistical independence properties in the data. The other one is the score-based method. It evaluates the quality of candidate causal models with some score functions and outputs one or multiple graphs having the optimal score \cite{GES, Heckerman95}. To calculate such scores, current prevailing methods assume a particular model class to describe causal mechanisms and data distributions. Widely-used score functions include the BIC/MDL score \cite{BIC} and the BGe score \cite{BGe} for linear-Gaussian models and the BDeu/BDe score \cite{BDeu, BDe} for discrete data. 

More recently, another set of approaches is based on restricted functional causal models, which represent the effect as a function of the direct causes together with an independent noise term. Under appropriate assumptions, they are able to identify the whole causal model. More specifically, the causal direction implied by the restricted functional causal model is generically identifiable; for example, the independence between the noise and causes holds only in the true causal direction and is violated in the wrong direction. Examples of such restricted functional causal models include the linear, non-Gaussian, acyclic model (LiNGAM \cite{LiNGAM}), the additive noise model \cite{nonlinear_CD, nonlinear_CD_kun}, and the post-nonlinear causal model \cite{post_nonlinear}.

\section{Overview and Contributions}

In this paper, we hypothesize atypical causal connectivities between brain regions for individuals with ASD and aim to identify them for diagnosis from the resting-state fMRI data. To achieve this goal, we propose a two-step method for large-scale and cyclic causal discovery. Particularly, the proposed approach to causal discovery relies on constrained functional causal models, which represents the effect as a linear function of the direct causes and some additive noise term. It is able to recover the whole causal structure between brain regions and allows feedbacks. The learned causal structure, as well as the causal influence strength, provides us the path and effectiveness of information flow; i.e., whether region $A$ directly influences region $B$, and if there is a direct influence, how strong it is. 
With the recovered causal influence strength as candidate features, we further perform feature selection and classification for ASD diagnosis. 

We apply our methods to three datasets from the ABIDE. From experimental results, it shows that with causal connectivities, the out-of-sample diagnostic accuracy largely improves, compared with using correlation and partial correlation as measures. A closer examination shows that information flow starting from the superior front gyrus to other areas are largely reduced; the target areas are mostly concentrated in default mode network and the posterior part. Moreover, all enhanced information flows are from posterior to anterior or in local areas, while local reductions also exist. Overall, it shows that long-range influences have a larger proportion of reductions than local influences, while local influences have a larger proportion of increases than long-range influences. By examining the graph properties of brain causal structure, the group of ASD shows reduced small-worldness compared to the control group. % do we need to add results here?

The contribution of this paper is mainly two-fold.
\begin{itemize}
	\item We proposes a method to identify atypical causal connectivities of ASD from resting-state fMRI data. The proposed two-step method identifies causal structures from observational data--interventions or randomized experiments are not needed. To the best of our knowledge, this is the first method to reliably extract atypical causal connectivities for ASD diagnosis without doing randomized experiments.
	\item With the proposed causal discovery and classification approach, the out-of-sample diagnostic accuracy achieves up to 87\%, which is better than other state-of-the-art approaches. Our neuropathology findings are in concordance with other widely recognized findings. It also suggests other diagnostic biomarkers of ASD.
\end{itemize}
%~\\

The outline of this paper is as follows.
\begin{itemize}
	\item [Chapter 2:] This chapter first provides a brief literature review on current diagnostic approaches in ASD,  including behavior diagnosis, genetic testing, and neuroimaging-based diagnosis.
	\item [Chapter 3:] This chapter focuses on computational methods for causal discovery from fMRI data. We will first give a brief summary of recent developments in causal discovery. The limitations of current approaches for causal discovery from fMRI data thus motivate our proposed two-step method. A detailed description of two-step method for large-scale and cyclic causal discovery will be followed.
	\item [Chapter 4:] This chapter gives experimental results by applying the proposed two-step methods to ABIDE datasets. 
	\item [Chapter 5:] This chapter concludes the paper and gives some suggestions for future research.
	
\end{itemize}

\chapter{Current ASD Diagnosis}
In this chapter, I will discuss main techniques in current ASD diagnosis, including behavior diagnosis (Chapter \ref{Sec:Behavior}), genetic testing (Chapter \ref{Sec:Gene}), and neuroimaging-based diagnosis (Chapter \ref{Sec:Image}). Specifically, behavior diagnosis has been widely used in clinic, while genetic testing and neuroimaging-based diagnosis mostly remain in the research stage.  
%behavior, fMRI, genetic

\section{Behavior Diagnosis} \label{Sec:Behavior}
Clinically, ASD diagnosis is mainly based on behavior diagnosis. The most-widely used ASD diagnostic criteria are based on the fifth edition of the Diagnostic and Statistical Manual of Mental Disorders (DSM-5) \cite{DSM5}. From DSM-5, a subject is diagnosized as ASD by specially trained physicians and psychologists if he/she satisfies the following key points.
\begin{itemize}
	\item [A.] Persistent deficits in social communication and social interaction across multiple contexts, as manifested by the following, currently or by history (examples are illustrative, not exhaustive): 
	\begin{itemize}
		\item [1.] Deficits in social-emotional reciprocity, ranging, for example, from abnormal social approach and failure of normal back-and-forth conversation; to reduced sharing of interests, emotions, or affect; to failure to initiate or respond to social interactions.
		
		\item [2.]  Deficits in nonverbal communicative behaviors used for social interaction, ranging, for example, from poorly integrated verbal and nonverbal communication; to abnormalities in eye contact and body language or deficits in understanding and use of gestures; to a total lack of facial expressions and nonverbal communication.
		
		\item [3.] Deficits in developing, maintaining, and understanding relationships, ranging, for example, from difficulties adjusting behavior to suit various social contexts; to difficulties in sharing imaginative play or in making friends; to absence of interest in peers.
	\end{itemize}
	\item [B.] Restricted, repetitive patterns of behavior, interests, or activities, as manifested by at least two of the following, currently or by history (examples are illustrative, not exhaustive):
	\begin{itemize}
		\item [1.] Stereotyped or repetitive motor movements, use of objects, or speech (e.g., simple motor stereotypes, lining up toys or flipping objects, echolalia, idiosyncratic phrases).
		
		\item [2.] Insistence on sameness, inflexible adherence to routines, or ritualized patterns or verbal nonverbal behavior (e.g., extreme distress at small changes, difficulties with transitions, rigid thinking patterns, greeting rituals, need to take same route or eat food every day).
		
		\item [3.] Highly restricted, fixated interests that are abnormal in intensity or focus (e.g., strong attachment to or preoccupation with unusual objects, excessively circumscribed or perseverative interest).
		
		\item [4.] Hyper- or hypo-reactivity to sensory input or unusual interests in sensory aspects of the environment (e.g., apparent indifference to pain/temperature, adverse response to specific sounds or textures, excessive smelling or touching of objects, visual fascination with lights or movement).
	\end{itemize}
	
	\item [C.]  Symptoms must be present in the early developmental period (but may not become fully manifest until social demands exceed limited capacities, or may be masked by learned strategies in later life).
	
	\item [D.]  Symptoms cause clinically significant impairment in social, occupational, or other important areas of current functioning.
	
	\item [E.]  These disturbances are not better explained by intellectual disability (intellectual developmental disorder) or global developmental delay. Intellectual disability and autism spectrum disorder frequently co-occur; to make comorbid diagnoses of autism spectrum disorder and intellectual disability, social communication should be below that expected for general developmental level.
\end{itemize}

The main limitation of behavior assessment is that it is unable to find out the underlying biological bases behind the observed behavioral symptoms. Knowledge of such biological bases may help to find a proper treatment. For example, if we can identify genetic causes or pathophysiology of ASD, it may motivate the development of corresponding medical treatment.

\section{Genetic Testing} \label{Sec:Gene}

Genetic testing can find changes in a person's DNA associated with specific disorders or conditions. Researchers have found that ASD has a strong genetic basis \cite{Genetic_ASD, GeneTest}. With genetic testing, it may help us to find out the underlying genetic cause of a child's autism, which thus may guide him/her to the most comprehensive therapy and treatment.  

\section{ Neuroimaging-based Diagnosis} \label{Sec:Image}
In past two decades, the development of neuroimaging techniques provides us noninvasive ways to uncover brain structures, to measure whole brain activities, and then to infer neuropathology biomarkers of ASD. 
There have been a number of research studies on ASD diagnosis from neuroimaging data. Different types of features have been considered and extracted for diagnosis, e.g., anatomical structures, brain activities, and brain connections. The neuroimaging-based diagnosis still remains in the research stage. It has not been used in clinics.

\subsection{Anatomical Structures}
Anatomical structures of the brain are usually obtained from magnetic resonance imaging (MRI), which has been a staple for autism research for quite some time. To conclude whether anatomical structure of a particular area differs in the group of typical controls and that of ASD, a statistical test is usually performed.

According to the review \cite{Autism_review1}, the most consistent finding is that of accelerated brain volume growth in early childhood for individuals with ASD, reported to be about a 10\% increase in brain volume, which seems to peak around 2-4 years of age. 
Structural studies of grey and white matter suggest an abnormal developmental trajectory of brain growth, with evidence of poorly organized white matter, increased cortical thickness, and atypicalities in gyration patterns. Preliminary morphology studies in regional specificities also have reported reduced volume in thalamus \cite{ASD_thalamus} and amygdala \cite{ASD_amygdala1, ASD_amygdala2, ASD_amygdala3} compared to control group. The local cortical thickness measure has been shown to be correlated with symptom severity in ASD and has been used as a predictor for it \cite{ASD_thick}.

\subsection{Brain Activities}
There are several neuroimaging techniques which can measure neural activities directly or indirectly, including fMRI, EEG, MEG, and electrophysiological techniques. In the following, we focus on fMRI related research studies. 
The fMRI provides a noninvasive way to measure whole brain activities. It works by detecting the changes in blood oxygenation and flow that occur in response to neural activities; when a brain area is more active, it consumes more oxygen, and to meet this increased demand, blood flow increases to the active area. The fMRI has several significant advantages: it is noninvasive, making it safe for the subject; it has excellent spatial and good temporal resolution; it is easy for the experimenter to use. The attractions of fMRI have made it a popular tool for imaging brain functions. To determine whether there are significant differences in neural activities in two groups, a significance test is usually performed as well.

Preliminary functional studies focusing on regional brain activation in autism have reported abnormal activities in a diverse set of brain regions, e.g., reduced activation in the amygdala when processing facial expressions \cite{ASD_amygdala4, ASD_amygdala5}, reduced activation within frontal, parietal, and occipital regions during a spatial attention task \cite{ASD_attention1}.

\subsection{Brain Connectivities}
Recently, with the hypothesis that ASD is associated with atypical brain connectivities, several studies examine differences in brain connectivities between ASD and typical controls from fMRI data. 
For example, hypoconnectivity between amygdala and cortical regions \cite{ASD_amygdala6} and between amygdala and insula \cite{ASD_insula1} have been found in adolescents and adults with autism compared with healthy controls.  Altered functional connectivities from the thalamus to the cortex could lead to impaired information transmission from the sensory periphery to the cortex in individuals with ASD \cite{ASD_thalamus2}. Research examining developmental changes in large-scale network functional connectivity reported reduced connectivity between amygdala/subcortical network and default mode network in adolescents with autism, as well as increased connectivity within amygdala/subcortical network in children with autism \cite{ASD_amygdala7}. The brains of individuals with autism exhibit reduced long-range connectivity \cite{Marcel_Just2}.

How can we identify brain connectivities for ASD diagnosis from fMRI data? Basically, there are three methods which have been mostly used: correlation, partial correlation, and Granger causality.

\subsubsection{Correlation}
Most work in the literature makes use of Pearson's correlation to learn functional connectivities between ROIs. Suppose that there are $n$ subjects, $m$ ROIs for each subject, and $T$ data points for each ROI and subject. We denote data from subject $r$ as $\mathbf{x}^{(r)}$, where $\mathbf{x}^{(r)}$ is a $m \times T$ matrix. The Pearson's correlation between ROI $i$ and ROI $j$ for subject $r$ is defined as
\begin{equation}
\rho_{ij}^{(r)} = \frac{\sum_{t= 1}^{T} (x_{it}^{(r)} - \bar{x}_{i:}^{(r)}) (x_{jt}^{(r)}- \bar{x}_{j:}^{(r)})}{\sqrt{\sum_{t= 1}^{T} (x_{it}^{(r)} - \bar{x}_{i:}^{(r)})^2} \sqrt{\sum_{t= 1}^{T}(x_{jt}^{(r)} - \bar{x}_{j:}^{(r)})^2}}
\end{equation}
where $x_{it}^{(r)}$ indicates the corresponding entry (the $i$th row, the $t$th column) of $\mathbf{x}^{(r)}$, ${x}_{i:}^{(r)}$ represents the $i$th row of $\mathbf{x}^{(r)}$, and $\bar{x}_{i:}^{(r)}= \frac{1}{T} \sum_{t= 1}^{T} x_{it}^{(r)}$; the similar notation is used for $\bar{x}_{j:}^{(r)}$. 

To determine whether two ROIs have a functional connectivity, we can either set a threshold on $\rho_{ij}^{(r)}$ or do statistical test. Let $B^{(r)}$ represent the learned connectivity matrix for subject $r$. Particularly, if we set a threshold $\alpha$ on $\rho_{ij}^{(r)}$, then $B_{ij}^{(r)} = 1$ if $\rho_{ij}^{(r)}>\alpha$; otherwise, $B_{ij}^{(r)} = 0$. If we do a statistical test, one may perform one-sample t-test to calculate the $p$ value. If the $p$ value is smaller than the significance level, then we reject the null hypothesis, and set $B_{ij}^{(r)} = 1$; otherwise, $B_{ij}^{(r)} = 0$.

After estimating the connectivity matrix $B_{ij}^{(r)}$ for each subject, one can feed the learned features to classification algorithms to do model estimation and prediction.

%With Pearson's correlation, it usually gives dense connections. By introducing an $L_1$ regularizer on it \cite{PearsonCorr_L1}:
%\begin{equation}
%\hat{\rho}_{ij} = \arg \min_{\rho_{ij}} \frac{1}{T} \sum_{t = 1}^{T}(x_{it} -\rho_{ij} x_{jt} )^2 + \lambda \sum_{i \neq j}|\rho_{ij}|,
%\end{equation}
%we can obtain a sparse solution.

\subsubsection{Partial Correlation}
There are studies using partial correlation to estimate functional connectivity between ROIs \cite{ParCorr1,ParCorr2}.
%The partial correlation test assumes that the underlying causal relation is linear and the data distribution is jointly Gaussian. 
Partial correlation is a measure of the strength of a linear relationship between two continuous variables whilst controlling for the effect of one or more other variables. Suppose that we have variables $X$ and $Y$, and a set of variables denoted as $\textbf{Z}$, the partial correlation $X$ and $Y$ given $\textbf{Z}$ is
\begin{equation*}
\hat{r}_{XY \cdot \textbf{Z}} = \frac{N \sum_{i=1}^{N} e_{X}(i) e_{Y}(i) - \sum_{i=1}^{N} e_{X}(i) \sum_{i=1}^{N} e_{Y}(i)}{\sqrt{N \sum_{i=1}^{N} {e_{X}^2(i)} - (\sum_{i=1}^{N} e_{X}(i))^2} \sqrt{N \sum_{i=1}^{N} {e_{Y}^2(i)} - (\sum_{i=1}^{N} e_{Y}(i))^2}},
\end{equation*}
where  $e_{X}(i)$ is the $i$th value of $e_{X}$, and $e_{X}$ is the residue after fitting a linear regression model of $X$ on $\mathbf{Z}$; similar notations are applied to $e_{X}(i)$.

Alternatively, we can estimate the partial correlation between every pair of ROIs giving all the remaining ROIs by learning the inverse covariance matrix. Let $\Sigma$ represent the covariance matrix between $m$ ROIs,  $\Theta = \Sigma^{-1}$, and $S$ be the empirical covariance matrix. The problem is to maximize the penalized log-likelihood \cite{InverseCov1}
\begin{equation}
\log \det \Theta - \mathtt{tr} (S \Theta) - \lambda \parallel \Theta \parallel_1
\end{equation}
over non-negative definite matrix $\Theta$, where $\mathtt{tr}$ denotes the trace, and $ \parallel \Theta \parallel_1$ is the $L_1$ norm. It is estimated by using a coordinate descent procedure.

\subsubsection{Granger Causality}
Granger causality has been used for the purpose of estimating the effective connectivity, which refers explicitly to the influence that one neural system exerts over another. Particularly, the connections between underlying neurons are learned by building a state space model to model the relations between observed fMRI signal $x$ and underlying neuronal activities $h$, and the causal connectivities between neuronal activities \cite{Autism_Granger1}. A dynamic state-space model can be described as follows.
\begin{equation}
\tilde{h}(t) = 
\begin{bmatrix}
h(t) \\
u(t) \\
\theta(t)	  
\end{bmatrix}
= 
\begin{bmatrix}
f(h(t-1), u(t-1),\theta(t-1)) \\
u(t-1)\\
\theta(t-1)
\end{bmatrix}
+
\begin{bmatrix}
P(t-1) \\
Q(t-1)\\
R(t-1)
\end{bmatrix}
\end{equation}
where $h$ is the hidden neuronal state variable, $u$ is the exogenous input, $\theta$ represents the hemodynamic response function (HRF) parameter variables. $f$ is the function links the current neuronal state to the previous neuronal states, exogenous inputs and parameters. $P$, $Q$, and $R$ are the zero mean Gaussian state noise vectors. The observation equation links the state to observation variables as given below.
\begin{equation}
x(t) = m(\tilde{h}(t)) + \epsilon(t-1)
\end{equation}
where $\epsilon$ is the measurement noise, and $m$ is the measurement function which links the state variables to measurement variables. The dimensionality of $x$ and $h$ is the same.

The hidden neuronal variables is learned by cubature Kalman filtering (CKF) \cite{CKF}. The estimated coefficients between neuronal activities are used as an input to a classification model. The reported diagnostic accuracy on 15 typical controls and 15 individuals with ASD from task-fMRI achieves up to 95.9\%. 

There are two points of this method that need to be investigated. The temporal resolution of neuronal activities and BOLD signals are quite different. The temporal resolution of neuron activities is at the scale of 1$\sim$100ms, while that of BOLD is around 2s. Roughly speaking, the BOLD signal can be seen as an aggregation of neural activities \cite{Gong_Aggre}. Thus, the identifiability of the state space model may not hold. In addition, the spatial resolution of neuronal activities and BOLD signal are also different. One BOLD signal may be the weighted average of activities from several neurons. Thus, it needs more careful considerations to build a one-to-one mapping between neuronal time series and fMRI time series. With the above problems, the reported accuracy achieved by this approach may be biased. It is important to test this method on other datasets to confirm its validity.

%~\\
After the estimation of brain connectivities, a proper classifier then is applied for diagnosis. The mostly widely-used classifiers include logistic regression, support vector machine \cite{SVM}, and random forests \cite{Random_forest}.

\chapter{The Two-Step Method for Cyclic and Large-Scale Causal Discovery}
	This chapter will focus on computational methods for causal discovery from fMRI data.
	In Chapter \ref{Sec: Motivation}, I will first give a brief summary of recent developments in causal discovery. Then I will discuss the limitations of these approaches for causal discovery from fMRI data, which thus motivate our proposed two-step method. Next in Chapter \ref{two-step}, I introduce the two-step method for causal discovery in detail, including model definition and model estimation. The identifiability of the two-step method will be discussed in Chapter \ref{Sec:Identifiability}.

	\section{Background Knowledge and Motivation} \label{Sec: Motivation}
	
	Identification of causal relationships is a fundamental problem in neural science. Traditionally, interventions or randomized experiments are used for inferring causal relationships. However, conducting such experiments is often expensive or even impossible, and from their results it is not easy to construct quantitative causal models. Alternatively, one may perform causal discovery from passively observational data.
	
	What kind of information can help to identify causal relationships? Correlation is useful to indicate a predictive relationship. However, it is not sufficient to demonstrate the presence of causation. The famous phrase ``Correlation does not imply causation" emphasizes that a correlation between two variables does not necessarily imply that one causes the other. Reichenbach's Common Cause Principle \cite{Reichenbach} tries to link causality and statistical dependence. It says that if $X$ and $Y$ are statistical dependent, then there are three possible relations between $X$ and $Y$: (1) $X$ causes $Y$. (2) $Y$ causes $X$. (3) A confounder $Z$ influences both $X$ and $Y$. There are other scenarios that may lead to the statistical dependence, such as selection bias on $X$ and $Y$ \cite{Zhang_selebias}, nonstationarity \cite{Huang15, Zhang_nonsta1, Huang17_ICDM}, and coincidence. %, while there are no causal relations between them. 
	
	Then how can we infer the causal structure from observational data in a principled way? The past few decades have seen the development of causal discovery.
	Roughly speaking, there are four types of methods.
	\begin{itemize}
		\item \textbf{Granger causality}. Granger causality \cite{Granger} is a statistical concept of causality based on prediction. It relies on two principles: (1) The cause happens prior to its effect. (2) The cause has unique information about the future values of its effect.
		
		According to Granger causality, if a signal $X_1$ ``Granger-causes"  a signal $X_2$, then past values of $X_1$ should contain information that helps predict $X_2$ above and beyond the information contained in past values of $X_2$ alone. %Its mathematical formulation is based on linear regression modeling of stochastic processes (Granger 1969). More complex extensions to nonlinear cases exist.
		
		\item \textbf{Constraint-based methods}. Based on some assumptions to relate statistical properties of the data and the causal structure, constraint-based methods use statistical tests (conditional independence tests) to find the causal skeleton and determine the orientations of edges up to the Markov equivalence class; all members of such a class share the same conditional independence relationships. The required assumptions include the causal Markov condition and the faithfulness assumption, which entail a correspondence between separation properties in the underlying causal structure and statistical independence properties in the data. Usually the constraint-based method consists of two components: conditional independence test and search method. The PC algorithm \cite{SGS}, developed by Peter Spirtes and Clark Glymour, is the most-widely used constraint-based method. It efficiently searches for acyclic causal relations among a set of variables up to the Markov equivalence class.
		
		Extentions of the PC algorithm include conservative PC (CPC) \cite{CPC}, fast causal inference (FCI) \cite{SGS}, and cyclic causal discovery (CCD) \cite{CCD}. The CPC algorithm requires a weaker assumption of faithfulness for asymptotically correct results, compared to PC. FCI allows for hidden common causes and selection bias; it recovers the causal graph up to an equivalence class in the presence of confounders -- partial ancestral graph (PAG). CCD is a constraint-based method which further considers cycles and outputs a PAG.
		
		\item \textbf{Score-based methods}. Score-based methods address causal structure learning as a model selection problem and are usually applied to learn a directed acyclic graph (DAG). They define a score function that measures how well a Bayesian network fits the observed data and searches through possible network structures to find the best scored causal network.  Similar to constraint-based methods, score-based methods have two components: proper score functions and search methods. 
		
		Current prevailing methods usually assume a particular model class to describe causal mechanisms and data distributions; widely-used score functions include the BIC/MDL score for linear-Gaussian models and the BDeu/BDe score for discrete data \cite{BDeu, BDe}. A generalized score function \cite{Huang_score} has been proposed recently, which does not assume a particular model class to describe causal mechanisms or data distributions. It can deal with a wide range of nonlinear causal relations and a wide class of data distributions, including non-Gaussian data and mixed continuous and discrete data.  It also applies to variables with different dimensions. Their proposed generalized score functions is based on the characterization of general (conditional) independence relationships between random variables, making use of kernels. One can use either cross-validated likelihood or marginal likelihood as score functions, both of which have shown to be statistically consistent in model selection under mild assumptions.
		
		One of the most popular search methods is the greedy equivalence search (GES) algorithm \cite{GES}. It searches through the space of Markov equivalence class and outputs a causal Markov equivalence class. It constitutes two phases: the forward phase and the backward phase. In the forward phase, it adds an edge greedily in each step, which maximizes the score function. It has been shown that all independence constraints in the resulting equivalence class from the first phase hold in the true data distribution. In the backward phase, it deletes edges also in a greedy manner. Asymptotically, it guarantees to find the Markov equivalence class which is consistent to the data generative distribution. It has shown to be more computational efficient compared to greedy search hill climbing, which searches through the DAG space.
		
		\item \textbf{Functional causal models}. 
		The functional causal model (FCM)-based approach is a statistical technique for estimating and testing causal relations using a combination of statistical data and qualitative causal assumptions. For each variable $X_i$, $i = 1, \cdots, m$, in a DAG $\mathcal{G}$, a FCM can be represented as \cite{Pearl}
		\begin{equation*}
		X_i = f(\mathbf{PA}_i^{\mathcal{G}}, E_i),
		\end{equation*}
		with independent noise terms $p(E_1,\cdots,E_m) = \prod_{i=1}^m p(E_i)$. 
		
		Recent developments in causal discovery have shown that by properly constraining the model class, the causal direction can be identifiable; that is, when we estimate the FCM in the right causal direction, the estimated noise term is independent of the hypothetical causes, but the independence does not hold in the anti-causal direction. Constrained functional causal models, which have shown to be identifiable, include the linear non-Gaussian acyclic model (LiNGAM) \cite{LiNGAM}, the nonlinear additive noise model \cite{nonlinear_CD_kun, nonlinear_CD}, and the post-nonlinear model \cite{post_nonlinear}.
		
	\end{itemize}
	
	Usually we assume that the data are independent, identically distributed (i.i.d.). More recently, it has been shown that distribution shifts (e.g. the nonstationarity) provide further benefits for causal direction identification \cite{Zhang_nonsta1}, which can be regarded as a kind of soft intervention. 
	
\begin{figure}[htp!]
		\begin{center}
			\begin{center}
				\begin{tikzpicture}[scale=.6, line width=0.5pt, inner sep=0.2mm, shorten >=.1pt, shorten <=.1pt]
				\draw (0, 0) node(1) [circle, draw] {{\footnotesize\,$X$\,}};
				\draw (2.5, 0) node(2) [circle, draw] {{\footnotesize\,$Y$\,}};
				\draw (0,-1.8) node(3) [circle, draw] {{\footnotesize\,$Z$\,}};
				\draw (2.5,-1.8) node(4) [circle, draw] {{\footnotesize\,$W$\,}};
				\draw[-latex] (1) -- (3); 
				\draw[-latex] (2) -- (4);
				\draw[-latex] (3.10)--(3.10-|4.west); %(3) -- (4); 
				\draw[-latex] (4.-170)--(4.-170-|3.east); %(4) -- (3);  
				
				\draw (4.5, 0) node(6) [circle, draw] {{\footnotesize\,$X$\,}};
				\draw (7, 0) node(7) [circle, draw] {{\footnotesize\,$Y$\,}};
				\draw (4.5,-1.8) node(8) [circle, draw] {{\footnotesize\,$Z$\,}};
				\draw (7,-1.8) node(9) [circle, draw] {{\footnotesize\,$W$\,}};
				\draw[-latex] (6) -- (9); 
				\draw[-latex] (7) -- (8);
				\draw[-latex] (8.10)--(8.10-|9.west); %(3) -- (4); 
				\draw[-latex] (9.-170)--(9.-170-|8.east); %(4) -- (3);
				\end{tikzpicture} \\
				(a)  ~~~~~~~~~~~~~~~~~~~~~~~(b) 
			\end{center}
			\caption{The two causal graphs are in the same PAG, which share the same conditional independence. They have different causal skeletons and causal directions.}
			\label{fig:PAG}
		\end{center}
	\end{figure}
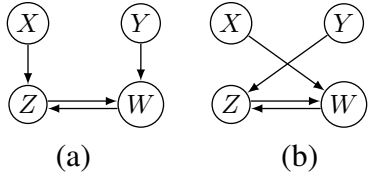

	%~\\
	%~\\
	Most methods in causal discovery, e.g., PC, all score-based methods, LiNGAM, nonlinear additive noise model, and post-nonlinear model, assume that the underlying causal graph is acyclic. Two methods, which can handle feedback loops, are CPC and linear non-Gaussian model (LiNGM) \cite{LiNGM}. However, they have their own limitations. The CPC algorithm makes use of conditional independence tests to recover the causal graph up to an equivalence class--the partial ancestral graph (PAG). Graphs in a PAG share the same conditional independence, but they may have different skeletons and orientations (Figure \ref{fig:PAG}). LiNGM belongs to the constrained functional causal model. Under the assumptions that the cycles are disjoint and that there are no self loops, the recovered stable graph is unique. The limitation of LiNGM is that it only works well on low-dimensional settings, especially when the sample size is small. A recent work \cite{Ruben19_fMRI} compares most causal discovery algorithms on simulated fMRI data.

	Brain signals from fMRI have the following characteristics. First, the connections between brain signals usually have feedback loops.  Second, the spatial resolution of fMRI signal is usually high, while the temporal resolution is relatively low. The high spatial resolution leads to a large set of variables for analysis, which increases computational complexity and difficulty in causal discovery. The low temporal resolution, as well as limited scanning period, results in limited data points which are available for analysis. The high dimensionality and limited sample size lead to the ``high-dimensional problem", which raises statistical challenges for estimation. The low temporal resolution also reduces time-lagged influences and increases instantaneous influences between brain signals \cite{Aggre_Gong}.
	
	Therefore, it is necessary to develop new causal discovery approaches, which are suitable for fMRI data analysis. Particularly, it should be able to handle feedback loops, and it should be computationally efficient and statistical reliable for high-dimensional settings with limited sample size.
	
	We propose a two-step method to learn the causal structure between regions of interests (ROIs) for each subject. The proposed approach is able to recover the whole causal structure reliably, including feedbacks among brain regions. It contains two steps. In the first step, we learn a superset of the underlying causal skeleton. In the second step, we identify the causal influence strength between ROIs for each subject. Particularly, the second step relies on the constrained functional causal model, which represents the effect as a function of the direct causes together with an independent noise term. After certain transformations, it reduces to an estimation of independent component analysis (ICA). The resulting causal structure is uniquely identifiable under certain conditions. The results from the first step are used as constraints of connections in the second step, which makes the estimation faster and more reliable, especially in the high-dimensional case.

	\section{Model Definition and Estimation} \label{two-step}
	
	As the name suggests, the two-step method for cyclic and large-scale causal discovery contains two steps. In the first step, it learns a superset of the causal skeleton, which is used as constraints of connections for the second step. In the second step, it uniquely identifies the causal structure, as well as causal strength, in a computationally efficient and statistically reliable way.  
	
	\subsection{First Step: Learning a Superset of Causal Skeleton} \label{Subsubsec: first step}
	In the first step, we learn a superset of the causal skeleton between ROIs. We denote by $O$ the learned causal skeleton from the first step, where $O_{jk} =1$ means that ROI $k$ is adjacent to ROI $j$ from the first step; otherwise, $O_{jk} = 0$. $O$ is symmetric. For computational efficiency, we let all subjects share the same adjacent matrix $O$ from the first step.
	
	Suppose that we have $n$ subjects in total, from the group of TCs and that of ASD. Further suppose that there are $m$ ROIs for each subject, and that there are $T$ data points per ROI per subject. Let the data for each subject be represented as $\mathbf{x}^{(i)}$ ($i=1, \cdots, n$), which is a $ m \times T$ matrix. We concatenate the data from all subjects and denote it as $\tilde{\mathbf{x}} = [\mathbf{x}^{(1)}, \cdots,\mathbf{x}^{(n)} ] $. Let $\tilde{x}_j$ ($j=1,\cdots,m$) denote the data from the $j$th ROI, i.e. the $j$th row of $\tilde{\mathbf{x}}$. To learn the superset of causal skeleton, we apply adaptive lasso regression \cite{Ada_lasso, ICA_sparse2} on each ROI $\tilde{x}_j$ ($j=1,\cdots,m$), with objective function as
	\begin{equation} \label{ada_lasso}
	\hat{\beta}_{jk}  = \arg \min_{\beta_{jk}} \big \{ \parallel \tilde{x}_j - \sum_{k \neq j } \beta_{jk} \tilde{x}_k \parallel_2^2 + \frac{\log n T}{2} \cdot \lambda \sum_{k \neq j} | \beta_{jk}/\check{\beta}_{jk} | \big \},
	\end{equation}
	where $\beta_{jk}$ is the regression coefficient, $\lambda$ is the parameter in the regularization term to control the sparsity of $\beta$, and $\check{\beta}_{jk}$ is a consistent estimator of $\beta_{jk}$. The consistent estimator $\check{\beta}_{jk}$ is estimated by Least Mean Squares regression
	\begin{equation*}
	\check{\beta}_{jk} = \arg \min_{\beta_{jk}} \big \{ \parallel \tilde{x}_j - \sum_{k \neq j } \beta_{jk} \tilde{x}_k \parallel_2^2  \big \}.
	\end{equation*}
	The adaptive lasso is used to overcome the disadvantage of lasso regression, including possible bias in the estimation of significant parameters. It has been shown that under mild conditions, the adaptive lasso estimation is consistent in model selection \cite{Ada_lasso}. After deriving  $\check{\beta}_{jk}$ from lasso regression, we apply it to (\ref{ada_lasso}) to estimate $\hat{\beta}_{jk}$.
	
	We set a threshold $\alpha$ on $\hat{\beta}_{jk}$. If $\hat{\beta}_{jk}$ is above the threshold, then $O_{jk}=1$; otherwise, $O_{jk}=0$. Finally, we symmetrize $O$: set $O_{jk}=1$, if $O_{kj}=1$, $\forall$ $j, k$. The diagonal entries of $O$ are constrained to be zero, which excludes self loops. 
	
	The learned superset of causal skeleton $O$ removes a large number of edges which do not exist in any subjects. It is used as a mask of possible edges in the second step; we fix those connections for all subjects to be zero if corresponding entries in $O$ are zero. It helps to make the estimation in the second step faster and more reliable, especially when the dimensionality is high.

	\subsection{Second Step: Identification of Causal Influences}
	Following from the notation in Chapter \ref{Subsubsec: first step}, we further denote the data for the $i$th subject and the $j$th ROI as $x_{j}^{(i)}$ ($i=1, \cdots, n, j=1,\cdots,m$) and denote its direct causes as $\mathbf{pa}_j^{(i)}$. Suppose that $x_{j}^{(i)}$ has the following data generating process
	\begin{equation} \label{FCM}
	x_j^{(i)} = \sum_{k} b_{jk}^{(i)} \mathbf{pa}_{j,k}^{(i)} + \mu_j^{(i)},
	\end{equation}
	where $\mathbf{pa}_{j,k}^{(i)} $ is the $k$th direct cause of $x_j^{(i)} $ with $\mathbf{pa}_j^{(i)} = \{\mathbf{pa}_{j,k}^{(i)} \}_k$, $b_{jk}^{(i)}$ represents corresponding causal coefficients, and $\mu_j^{(i)}$ represents the disturbance term due to omitted factors, which are assumed to be non-Gaussian distributed and are independent of each other.% In addition, the noise term $\mu_j^{(i)}$ is independent of the causes $\mathbf{pa}_{j}^{(i)}$.
	
	By considering all ROIs simultaneously, equation (\ref{FCM}) can be represented in the matrix form:
	\begin{equation}\label{FCM_matrix}
	\mathbf{x}^{(i)} = B^{(i)} \mathbf{x}^{(i)} + U^{(i)},
	\end{equation}   
	where $B^{(i)} = \{b_{jk}^{(i)}\}_{j,k}$ and $U^{(i)} = \{\mu_j^{(i)}\}_j$. By moving $\mathbf{x}^{(i)}$ to the left side, we have $\mathbf{x}^{(i)} = (I-B^{(i)})^{-1} U^{(i)}$. Since all terms in $U^{(i)}$ are non-Gaussian distributed and are independent of each other, it turns to be an ICA-based problem \cite{LiNGAM, LiNGM}. Let $W^{(i)} = I-B^{(i)}$. 
	
	\paragraph{Model estimation}
	We estimate the demixing matrix $W^{(i)}$ by maximizing the following penalized log-likelihood based on adaptive lasso
   \begin{equation}\label{ICA_obj}
	\mathcal{L}_{\lambda} = \sum_{i=1}^{n} \sum_{t=1}^{T} \sum_{j=1}^{m} \log p_j^{(i)} (\mathbf{w}_j^{(i) T}  \mathbf{x}^{(i)}(t) ) + T \log |W^{(i)}| - \lambda  \sum_{i, j, k} |W_{jk}^{(i)}/\check{W}_{jk}^{(i)}|,
	\end{equation}
	with the log-likelihood \cite{ICA_book}
	\begin{equation}
	\mathcal{L} =  \sum_{i=1}^{n} \sum_{t=1}^{T} \sum_{j=1}^{m} \log p_j^{(i)} (\mathbf{w}_j^{(i) T}  \mathbf{x}^{(i)}(t) ) + T \log |W^{(i)}|,
	\end{equation}
	where  $\mathbf{w}_j^{(i) T}$ is the $j$th row of $W^{(i)}$, $p_j^{(i)}(\cdot)$ is the probability density function, $\mathbf{x}^{(i)}(t)$ is the $t$th data point of $\mathbf{x}^{(i)}$, $ \lambda \sum_{i, j, k} |W_{jk}^{(i)}/\check{W}_{jk}^{(i)}|$ is the penalty term to enforce sparsity on $W^{(i)}$, and $\check{W}_{jk}^{(i)}$ is a consistent estimator of $W_{jk}^{(i)}$ estimated by
	\begin{equation*}
	 \check{W}_{jk}^{(i)} = \arg \max_{W_{jk}^{(i)}} \big\{\sum_{i=1}^{n} \sum_{t=1}^{T} \sum_{j=1}^{m} \log p_j^{(i)} (\mathbf{w}_j^{(i) T}  \mathbf{x}^{(i)}(t) ) + T \log |W^{(i)}| \big\}.
	\end{equation*}
	%$\sum\limits_{i, j, k} P_{\gamma}^{\text{GSCAD}}(W_{jk}^{(i)}) $ and $\sum\limits_{i_1 \neq i_2, j, k} P_{\lambda}^{\text{GSCAD}} (W_{jk}^{(i_1)}-W_{jk}^{(i_2)})$ are two regularization terms using generalized smoothly clipped absolute deviation (GSCAD) penalty \cite{ICA_sparse1}, constraining the sparsity of $W^{(i)}$ and that of $W^{(i_1)}-W^{(i_2)}$, respectively. By enforcing sparsity on $W^{(i)}$, it reduces model complexity and makes parameter estimation more reliable. In addition, the sparsity ensures that each ROI is influenced by a small number of causes, which may make the interpretation easier. By constraining the sparsity on the difference $W^{(i_1)}-W^{(i_2)}$, it ensures the similarity of connections across subjects. 
	
	In practice, we found that $W_{jk}^{(i)}$ are not reliably estimated, because the duration of fMRI scan for each subject is short, i.e., the sample size is small. To deal with this issue, we leverage all subjects to better estimate shared parameters, by constraining that the differences between subjects are small, which leads to much larger sample size. We modify the model estimation in (\ref{ICA_obj}) to 
	\begin{equation}\label{ICA_obj_updated}
      \tilde{\mathcal{L}}_{\lambda} = \mathcal{L} - \sum_{i, j, k} P_{\lambda}(W_{jk}^{(i)}) - \sum_{i_1 \neq i_2, j, k} P_{\gamma} (W_{jk}^{(i_1)}-W_{jk}^{(i_2)}),
    \end{equation}
    where $P_{\lambda}(W_{jk}^{(i)})$ and $P_{\gamma} (W_{jk}^{(i_1)}-W_{jk}^{(i_2)})$ are two regularization terms, with $P_{\lambda}(W_{jk}^{(i)}) = |W_{jk}^{(i)}|$ and $P_{\gamma} (W_{jk}^{(i_1)}-W_{jk}^{(i_2)}) = |W_{jk}^{(i_1)}-W_{jk}^{(i_2)}|$. By enforcing sparsity on $W^{(i)}$ and $W^{(i_1)}-W^{(i_2)}$, it reduces model complexity and makes parameter estimation more reliable. In addition, the sparsity on $W^{(i)}$ ensures that each ROI is influenced by a small number of causes. The sparsity on the difference $W^{(i_1)}-W^{(i_2)}$ ensures the similarity of connections across subjects. It has been shown that under the irrepresentable condition given in \cite{Lasso_consistency}, lasso regression is asymptotically consistent in model selection. 
	
	Since $|w|$ is not differentiable at the origin, we cannot directly use the gradient-based learning rule to optimize the objective function when $\omega$ is around zero. We use $tanh(m\omega)$ to approximate the derivative of $|\omega|$ \cite{ICA_sparse1}, where $m$ is a constant value. Then $P_{\lambda}(\omega)$ can be approximated by  $\lambda \cdot \tanh(m \omega)$. 
   
	The demixing matrix $W^{(i)}$ then can be updated by gradient descent, 
	\begin{equation}
	W^{(i)} = W^{(i)} + \eta \cdot \Delta W^{(i)}
	\end{equation}
	with  
	\begin{equation} 
	\Delta W^{(i)} = (W^{(i) T})^{-1} + E\{ g(W^{(i)} \mathbf{x}^{(i)}) \mathbf{x}^{(i) T}\} -  \sum\limits_{i} \mathbf{P}'_{\gamma} (|W^{(i)}|) - \sum\limits_{i_1 \neq i_2} \mathbf{P}'_{\lambda} (|W^{(i_1)} - W^{(i_2)}|),
	\end{equation} 
	where
	$ {\mathbf{P}'_{\gamma} (|W^{(i)}|)}_{jk} = P'_{\gamma} (|W_{jk}^{(i)}|) = \tanh (m |W_{jk}^{(i)}|)$, ${\mathbf{P}'_{\lambda} (|W^{(i_1)} - W^{(i_2)}|)}_{jk} = P'_{\gamma} (|W_{jk}^{(i_1)} - W_{jk}^{(i_2)}|) =  \tanh (m |W_{jk}^{(i_1)} - W_{jk}^{(i_2)}|) $,  	
	and $g = [g_1; \cdots; g_j; \cdots; g_m]$ is a function defined as
	\begin{equation*}
	g_j(\mu_j) = \frac{\partial}{\partial \mu_j} log(p(\mu_j)) = \frac{p'(\mu_j)}{p(\mu_j)},
	\end{equation*}
	with the density function $p(\cdot)$ estimated by kernel density estimation. 
	
	For those entries which are zero in $O$ derived from the first step, we fix the corresponding entries to be zero in $B^{(i)} = I-W^{(i)}$; that is, there are no direct causal connections between those variables. Denote $\tilde{O} = I - O$. We apply the mask $\tilde{O}$ to $\Delta W^{(i)}$ 
	\begin{equation*}
	\Delta \tilde{W}^{(i)} = \tilde{O} \odot \Delta W^{(i)}
	\end{equation*}
	and to $W^{(i)}$
	\begin{equation*}
	\tilde{W}^{(i)} = \tilde{O} \odot W^{(i)} 
	\end{equation*}
	to guarantee corresponding entries to be constantly zero, where $\odot$ represents the dot product. Then the update rule of $\tilde{W}^{(i)}$ is
	\begin{equation}
	\tilde{W}^{(i)} = \tilde{W}^{(i)} + \eta \cdot \Delta \tilde{W}^{(i)}
	\end{equation}

	\paragraph{Initialization of $W^{(i)}$}
	To guarantee the stability of resulting estimation, the initial value of $W^{(i)}$ is set in the following way.
	\begin{itemize}
		\item Set $W_0 = \tilde{O} \odot (I - \beta)$.
		\item Perform ICA on $\tilde{\mathbf{x}}$ with adaptive lasso regularization, with the initial demixing matrix $W_0$. Denote the estimation as $\tilde{W}_0$.    The matrix $\tilde{W}_0$ is used as a initialization of $W^{(i)}$ ($i = 1,\cdots,n$).
	\end{itemize}
	
	Let $\hat{\tilde{W}}$ represent the estimation of $\tilde{W}$. The causal influence strength matrix for subject $i$ is $\hat{B}^{(i)} = I-\hat{\tilde{W}}$. Algorithm \ref{Algo_twostep} gives the procedure of the two-step method.

	\begin{algorithm}   	
		\caption{The two-step method for cyclic and large-scale causal discovery}
		\textbf{Input}: Observed data $\mathbf{x}^{(i)}$ ($i = 1,\cdots,n$) \\
		\textbf{Output}: Causal influence strength $B^{(i)}$ for each subject\\
		1. The first step: learning a superset of causal skeleton 
		\begin{algorithmic}
			\State $\check{\beta}$ = Least Mean Square regression ($\tilde{\mathbf{x}}$)
			\State $\hat{\beta}$ = Adaptive lasso regression($\tilde{\mathbf{x}}$, $\check{\beta}$)
			\State $O = (|\hat{\beta}| > \alpha)$
			\State $O= O + O^T$
			\State $O = (O \neq 0)$
		\end{algorithmic}	 
		2. The second step: identifying the causal influences
		\begin{algorithmic}	 
			\State $W_0 =  \tilde{O} \odot (I - \beta)$
			\State $\tilde{W}_0 = $ ICA with adaptive lasso ($\tilde{\mathbf{x}}$)
			\State Initialize $\tilde{W}^{(i)} = \tilde{W}_0 $, $ \Delta \mathbf{W} = 0$
			\While {$\parallel \Delta \mathbf{W}\parallel_1 > 1e-6$}
			\State $\Delta \mathbf{W}= 0$
			\For {$i = 1 \text{ to } n$}
			\State $\Delta \tilde{W}^{(i)} = \tilde{O} \odot \Delta W^{(i)} $
			\State $\tilde{W}^{(i)} = \tilde{W}^{(i)} + \eta \Delta \tilde{W}^{(i)}$
			\State $\Delta \mathbf{W} = \Delta \mathbf{W} + \Delta \tilde{W}^{(i)}$
			\EndFor
			\EndWhile
			\For {$i = 1 \text{ to } n$}
			\State $B^{(i)} = I - \tilde{W}^{(i)} $
			\EndFor
		\end{algorithmic}	
		\label{Algo_twostep}
	\end{algorithm}
	
	\section{Model Identification} \label{Sec:Identifiability}
	
	We say that a causal model is identifiable if the objective function $\mathcal{L}$ has a unique optimal at $W = W^*$ when the sample size is large enough, i.e., for any $W' \neq W^*$, $\mathcal{L}(W') < \mathcal{L}(W^*)$. If the model is identifiable, then we can uniquely identify the causal structure asymptotically, since there is one and only one underlying causal structure.
	
	\subsection{Identifiability of ICA} \label{subSec: Iden_ICA}
	In the second step of the two-step method, we make use of the ICA estimation, so we first give the identifiability condition of ICA.
	
	Let $X$ represent observed variables. For $X = AS$ with sources in $S$ being independent of each other, if one of the following two conditions are satisfied \cite{ICA_identifiability, ICA_Comon}:
	\begin{itemize}[noitemsep]%,nolistsep] 
		\item  there is no Gaussian sources in $S$;
		\item  $A$ is full column rank and at most one source is Gaussian;
	\end{itemize}
	then the mixing matrix $A$ can be identified up to scale and permutation.
	
	In our scenario, the mixing matrix $A = W^{-1}$ is full column rank, so we allow that at most one error term is Gaussian.

	\subsection{Identifiability of Causal Model with Two-Step Method}
	
	We will show that under certain assumptions the two-step method can uniquely identify the underlying causal structure, without scale and permutation ambiguity. Particularly, the first step provides a superset of underlying causal skeleton. We first show that from the first step all true causal edges are kept, but there may exist spurious edges. 
	
	In the first step, we concatenate the data from $n$ subjects and learn a adjacency matrix $O$ on the concatenated datasets. Theorem \ref{moral} shows that if we do adaptive lasso regression on each subject separately, the learned adjacency matrix $O^{(i)}$ for subject $i$ is the moral graph of its causal graph. We first give the definition of the moral graph on the directed cyclic graph.
	
	\begin{Definition}
		The moral graph $M(G)$ of a directed cyclic graph $G$ is an undirected graph that contains an undirected edge between variable $X$ and $Y$, if
		\begin{itemize}
			\item there is a directed edge between them in either direction,
			\item or $X$ and $Y$ are the parents of the same variable.
		\end{itemize}
	\end{Definition}
	
%	\begin{lemma} \label{moral}
%		Let $O^{(i)}$ represent the adjacency matrix by doing adaptive lasso regression on the $i$th subject, and let $G^{(i)}$ represent the causal graph of the $i$th subject. Then under the faithfulness assumption, $O^{(i)}$ is the moral graph of $G^{(i)}$.
%	\end{lemma}
    % Let $P_{jk}^{(i)} = -B_{jk}^{(i)} \sigma_j^{-2 (i)} -B_{kj}^{(i)} \sigma_k^{-2 (i)} + \sum_{r \neq j, r \neq k} B_{rj}^{(i)} B_{rk}^{(i)} \sigma_r^{-2 (i)}$.
	\begin{lemma} \label{moral}
		Let $O^{(i)}$ represent the adjacency matrix by doing adaptive lasso regression on the $i$th subject, $G^{(i)}$ the causal graph of the $i$th subject, and $U^{(i)}$ the noise terms in the functional causal model of the $i$th subject.  Let $\sigma_j^{2 (i)}$ be the variance of noise term of $j$th ROI , i.e., $U_j^{(i)}$. Then under the condition that 
		\begin{equation} \label{condition}
		 	\forall j \neq k, \text{if } B_{jk}^{(i)} \cdot B_{kj}^{(i)} \neq 0, -B_{jk}^{(i)} \sigma_j^{-2 (i)} -B_{kj}^{(i)} \sigma_k^{-2 (i)} + \sum_{r \neq j, r \neq k} B_{rj}^{(i)} B_{rk}^{(i)} \sigma_r^{-2 (i)} \neq 0,
		\end{equation}	
		$O^{(i)}$ is the moral graph of $G^{(i)}$.
	\end{lemma}

	\begin{proof}
		Since $X^{(i)} = (I-B^{(i)})^{-1} U^{(i)}$, the covariance matrix of $X^{(i)}$ can be represented as 
		\begin{equation*}
		 C_X^{(i)} = (I-B^{(i)})^{-1} C_U^{(i)} (I-B^{(i)})^{-T},
		\end{equation*}
		where $C_X^{(i)}$ and $C_U^{(i)}$ are the covariance matrices of $X^{(i)}$ and $U^{(i)}$, respectively. Then the inverse covariance matrix $P_X^{(i)}$ is
		\begin{equation*}
		P_X^{(i)} = C_X^{(i) -1} = (I-B^{(i)})^{T} C_U^{(i) -1} (I-B^{(i)}).
		\end{equation*}
		Since $U^{(i)}$ is independent with each other, $C_U^{(i)}$ is a diagonal matrix, so is $C_U^{(i) -1}$. %Let $A = C_U^{-1/2}(I-B)$. Then $P_X = A^T A$. It is easy to see that if $B_{ij} \neq 0$, $A_{ij} \neq 0$. 
		The $(j,k)$th entry of $P_X^{(i)}$ is 
		\begin{equation*}
		 P_{jk}^{(i)} = -B_{jk}^{(i)} \sigma_j^{-2 (i)} -B_{kj}^{(i)} \sigma_k^{-2 (i)} + \sum_{r \neq j, r \neq k} B_{rj}^{(i)} B_{rk}^{(i)} \sigma_r^{-2 (i)},
		\end{equation*}
		which represents the partial correlation between $X_j^{(i)}$ and $X_k^{(i)}$ given all remaining variables.
		
		%The condition given in (\ref{condition}) ensure that if there is an edge between node $j$ and $k$, then the precision matrix of corresponding entries $P_{jk}$ and $P_{kj}$ is not zero.
		
		Under the condition that if $B_{jk}^{(i)} \neq 0$ or $B_{kj}^{(i)} \neq 0$, then $P_{jk}^{(i)} \neq 0, \forall j \neq k$, with adaptive lasso regression, all variables in the Markov blanket of the target variable will be in the set of predictors with nonzero coefficients. Thus, all variables in its Markov blanket will have a connection with the target variable. 
	\end{proof}
	From Lemma \ref{moral}, we have $\mathcal{S}(G^{(i)}) \subseteq \mathcal{S}(O^{(i)})$ for $i = 1,\cdots,n$, where $\mathcal{S}(\cdot)$ represent the skeleton of corresponding graph. The following theorem shows that the skeleton of the union of the moral graph $O^{(i)}$ is a subset of $O$.     
	
	\begin{lemma} \label{union}
		Let $O$ be the adjacency matrix estimated from the first step, and let $O^{(i)}$ represent the adjacency matrix by doing adaptative lasso regression on the $i$th subject ($i = 1,\cdots,n$). Then $\cup_i \mathcal{S}(O^{(i)}) \subseteq \mathcal{S}(O)$.
	\end{lemma}
	
	\begin{theorem}
		Let $O$ be the adjacency matrix estimated from the first step. For any subject $i$, under the condition given in (\ref{condition}), its causal skeleton $\mathcal{S}(G^{(i)})$ is a subset of $\mathcal{S}(O)$.
	\end{theorem}
	
	\begin{proof}
		From Lemma \ref{moral}, it is easy to infer that $\cup_i \mathcal{S}(G^{(i)}) \subseteq \cup_i \mathcal{S}(O^{(i)})$. From Lemma \ref{union}, we have $\cup_i \mathcal{S}(O^{(i)}) \subseteq \mathcal{S}(O)$, and then $\cup_i \mathcal{S}(G^{(i)}) \subseteq \mathcal{S}(O)$. Therefore, $\mathcal{S}(G^{(i)}) \subseteq \mathcal{S}(O)$.
	\end{proof}

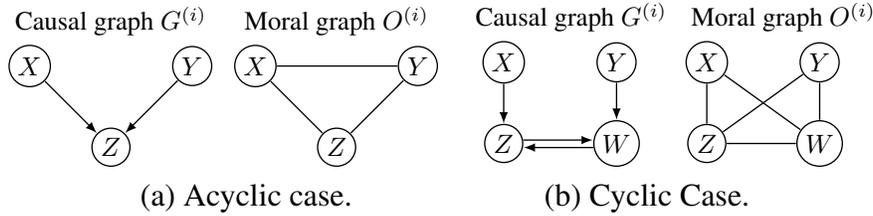
\begin{figure}[htp]
		\begin{center}
			\begin{center}
				\begin{tikzpicture}[scale=.6, line width=0.5pt, inner sep=0.2mm, shorten >=.1pt, shorten <=.1pt]
				\draw (0, 0) node(1) [circle, draw] {{\footnotesize\,$X$\,}};
				\draw (3.6, 0) node(2) [circle, draw] {{\footnotesize\,$Y$\,}};
				\draw (1.8,-1.8) node(3) [circle, draw] {{\footnotesize\,$Z$\,}};
				\draw[-latex] (1) -- (3); 
				\draw[-latex] (2) -- (3); 
				\draw (1.8, 1) node(7) [] {{\footnotesize\,Causal graph $G^{(i)}$\,}};
				
				\draw (5, 0) node(4) [circle, draw] {{\footnotesize\,$X$\,}};
				\draw (8.6, 0) node(5) [circle, draw] {{\footnotesize\,$Y$\,}};
				\draw (6.8,-1.8) node(6) [circle, draw] {{\footnotesize\,$Z$\,}};
				\draw[-] (4) -- (6); 
				\draw[-] (5) -- (6); 
				\draw[-] (4) -- (5); 
				\draw (6.8, 1) node(8) [] {{\footnotesize\,Moral graph $O^{(i)}$\,}};
				\end{tikzpicture} ~~
				\begin{tikzpicture}[scale=.6, line width=0.5pt, inner sep=0.2mm, shorten >=.1pt, shorten <=.1pt]
				\draw (0, 0) node(1) [circle, draw] {{\footnotesize\,$X$\,}};
				\draw (2.5, 0) node(2) [circle, draw] {{\footnotesize\,$Y$\,}};
				\draw (0,-1.8) node(3) [circle, draw] {{\footnotesize\,$Z$\,}};
				\draw (2.5,-1.8) node(4) [circle, draw] {{\footnotesize\,$W$\,}};
				\draw[-latex] (1) -- (3); 
				\draw[-latex] (2) -- (4);
				\draw[-latex] (3.10)--(3.10-|4.west); %(3) -- (4); 
				\draw[-latex] (4.-170)--(4.-170-|3.east); %(4) -- (3);  
				\draw (1.5, 1) node(5) [] {{\footnotesize\,Causal graph $G^{(i)}$\,}};
				
				\draw (4.5, 0) node(6) [circle, draw] {{\footnotesize\,$X$\,}};
				\draw (7, 0) node(7) [circle, draw] {{\footnotesize\,$Y$\,}};
				\draw (4.5,-1.8) node(8) [circle, draw] {{\footnotesize\,$Z$\,}};
				\draw (7,-1.8) node(9) [circle, draw] {{\footnotesize\,$W$\,}};
				\draw[-] (6) -- (8); 
				\draw[-] (7) -- (9);
				\draw[-] (8) -- (9); 
				\draw[-] (6) -- (9);  
				\draw[-] (7) -- (8); 
				\draw (6.2, 1) node(10) [] {{\footnotesize\,Moral graph $O^{(i)}$\,}};
				\end{tikzpicture} \\
				(a) Acyclic case. ~~~~~~~~~~~~~~~~~~~~~~~(b) Cyclic Case.
			\end{center}
			\caption{Moralization of causal graph. (a) Acyclic case.  (b) Cyclic case. }
			\label{fig:moral}
		\end{center}
	\end{figure}

\begin{figure}[htp]
		\setlength{\abovecaptionskip}{0pt}
		\setlength{\belowcaptionskip}{0.5pt}
		\begin{center}
			\setlength{\abovecaptionskip}{-0.2pt}
			\setlength{\belowcaptionskip}{0pt}
			\begin{center}
				\begin{tikzpicture}[scale=.6, line width=0.5pt, inner sep=0.2mm, shorten >=.1pt, shorten <=.1pt]
				\draw (0, 0) node(1) [circle, draw] {{\footnotesize\,$X$\,}};
				\draw (1.8, 0) node(2) [circle, draw] {{\footnotesize\,$Y$\,}};
				\draw (3.6, 0) node(3) [circle, draw] {{\footnotesize\,$Z$\,}};
				\draw (5.4, 0) node(4) [circle, draw] {{\footnotesize\,$W$\,}};
				\draw (3.6, 1.3) node(5) {{\footnotesize\,{\small$g(C)$}\,}};
				%  \draw (1.7,1.2) node(4) [circle, draw] {{\large\,$\psi$\,}};
				\draw[-latex] (1) -- (2); %node[pos=.5,above] {{\large$a_1$}};
				\draw[-latex] (2) -- (3); %node[pos=.5,above] {{\large$a_2$}};
				\draw[-latex] (3) -- (4); %node[pos=.5,above] {{\large$a_2$}};
				\draw[-latex] (5) -- (2); %node[pos=.5,above] {{\large$a_2$}};
				\draw[-latex] (5) -- (4); %node[pos=.5,above] {{\large$a_2$}};
				%  \draw[-arcsq] (4) -- (3);
				\end{tikzpicture} ~~
				\begin{tikzpicture}[scale=.6, line width=0.5pt, inner sep=0.2mm, shorten >=.1pt, shorten <=.1pt]
				\draw (0, 0) node(1) [circle, draw] {{\footnotesize\,$X$\,}};
				\draw (1.8, 0) node(2) [circle, draw] {{\footnotesize\,$Y$\,}};
				\draw (3.6, 0) node(3) [circle, draw] {{\footnotesize\,$Z$\,}};
				\draw (5.4, 0) node(4) [circle, draw] {{\footnotesize\,$W$\,}};
				%  \draw (1.7,1.2) node(4) [circle, draw] {{\large\,$\psi$\,}};
				\draw[-] (1) -- (2); %node[pos=.5,above] {{\large$a_1$}};
				\draw[-] (2) -- (3); %node[pos=.5,above] {{\large$a_2$}};
				\draw[-] (3) -- (4); %node[pos=.5,above] {{\large$a_2$}};
				\draw[-] (1) to[out=30,in=150] (4); %node[pos=.5,above] {{\large$a_2$}};
				\draw[-] (2) to[out=-30,in=-150] (4); %node[pos=.5,above] {{\large$a_2$}};
				%  \draw[-arcsq] (4) -- (3);
				\end{tikzpicture} \\
				~(a) ~~~~~~~~~~~~~~~~~~~~~~~~~~~~~~(b)
			\end{center}
			\vspace{.4cm}
			\caption{An illustration on how changes in the causal model may lead to spurious edges; $g(C)$ indicates that corresponding causal modules change across subjects. (a) The true causal graph (including confounder $g(C)$). (b) The estimated conditional independence graph on the observed data in the asymptotic case. }
			\label{fig:nonSta}
		\end{center}
		%  \vspace{1pt}
		%%\end{wrapfigure}
	\end{figure}
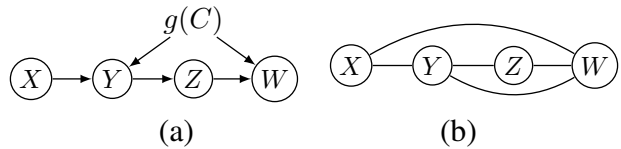

	The estimation from the first step gives a superset of the causal skeleton. We summarize possible sources of spurious edges.
	\begin{itemize}
		\item [1.] From the moralization of the casual structure: \\
		If $X$ and $Y$ are the parents of the same node in $G^{(i)}$, and they are not directly connected. However, in the moral graph $O^{(i)}$, $X$ and $Y$ are connected, and so is in $O$. Figure \ref{fig:moral} gives an illustration.
		%Due to moral graph (define moral graph of a cyclic graph) (prove with d-separation) ("Identifying independies in causal graphs with feedback")
		\item [2.] From the inconsistent causal structure across subjects: \\
		Suppose that a subset of subjects (denoted as $\mathcal{I}_1$) have causal edges from ROI $j$ to ROI $k$, while others (denoted as $\mathcal{I}_2$) do not. Our method will include the edge $i - j$ in $O$, which is spurious for those in $\mathcal{I}_2$.
		\item [3.] From the distribution shift across subjects: \\
		The spurious edges may result from heterogeneity, e.g. in the case when data from different subjects have different probability distributions and the changes of different ROIs are dependent. If concatenating different subjects, it may lead to spurious edges \cite{Zhang_nonsta1}. Figure \ref{fig:nonSta} gives an illustration.
	\end{itemize}

	Next, we are going to show that the second step can uniquely identify the causal structure, if the underlying causal model is stable and the cycles are disjoint. A general way to express stability is $\lim\limits_{t\rightarrow \infty} B^t = 0$, where $B$ is the causal influence matrix; it is mathematically equivalent to: for all eigenvalues $e$ of $B$, $|e|<1$ \cite{LiNGM}. We first show that without the stability and disjoint-cycle assumptions, the causal model is identifiable up to the population distribution; that is, all causal models entail the same probability distribution.
	\begin{lemma} \label{LiNGM1}
		If the underlying causal relations are linear, and the error terms are jointly independent with at most one being Gaussian, then under the condition given in (\ref{condition}), the two-step method outputs causal models that entail the probability distribution in the large sample limit.
	\end{lemma}   
	\begin{proof}
		We have shown that all true edges will be kept from the first step. In the second step, we make use of ICA estimation. ICA gives pointwise consistent estimation up to the population distribution \cite{ICA_Comon} under the assumptions listed in Chapter \ref{subSec: Iden_ICA}. Since adaptive lasso is unbiased, in the large sample limit, the estimation of the two-step method is pointwise consistent up to the population distribution. 
	\end{proof} 
	
	Now we show that by adding the stability and disjoint-cycle assumptions, the causal model is uniquely identifiable.        
	\begin{lemma}[\cite{LiNGM}] \label{LiNGM2}
		Suppose that a causal model $M$ with disjoint cycles entails a distribution $Q$, and that it is stable. Then for any other causal model $M'$ which entails the same distribution, $M'$ is not stable.
	\end{lemma}
	
	Lemma \ref{LiNGM2} infers that with the stability and disjoint-cycle conditions, causal model $M$ is identifiable; that is, there is one and only one causal model which satisfies the conditions. However, it should be noted that with ICA, the row sequence of $W$ is not determined; that is, the output directly from ICA may be not stable, and further permutation is required to get a stable solution. To achieve the goal that the output directly from ICA is stable, in the two-step method we ensure that the initial value of $W^{(i)}$ and its update in every iteration are stable. %Thus, with the two-step method, the output is stable; no further permutation is needed.

	\begin{theorem} \label{Identifiability_Twostep}
		Suppose that the underlying causal structure is stable. Further suppose that the cycles contained in the underlying graph are disjoint, and there are no self loops. Then  under the condition given in (\ref{condition}), with the two-step method, the causal model is identifiable, and the output from the two-step method directly gives the stable solution.
	\end{theorem}
	
	\begin{proof}
		From Lemma \ref{LiNGM1} and Lemma \ref{LiNGM2}, the two-step method is uniquely identifiable under the stable and disjoint-cycle conditions. By constraining that the initial value of $W^{(i)}$ and its update in every iteration is stable, the output from the two-step method directly gives the stable solution.
	\end{proof}

	\subsection{Discussion}
	The two-step method for cyclic and large scale causal discovery is computationally and statistically efficient.
	
	\textit{Cyclic causal discovery}: The two-step method is able to handle causal graph with cycles. By making use of the independence constraints that the error terms are independent of each other, the two-step method is able to handle cycles. With the assumption that the cycles are disjoint, then the stable solution is unique.
	
	\textit{Large-scale causal discovery}: The first step learns a superset of the causal structure, which acts as constraints of connections in the second step. The first step can be done in parallel, and thus, the first step is appropriate for large-scale problems. 
	The original ICA cannot handle large-scale problems, with the computational complexity $O(m(m+1)nT \cdot N)$, where $N$ is the number of iterations in the estimation. Fortunately, with the constraints from the first step, the number of parameters that need to be estimated in the second step are greatly reduced. Therefore, the two-step method is appropriate for large-scale problems.
	
	For the first step, alternatively, one can use fast adjacency search (FAS) \cite{SGS}, the first step of PC, to find a superset of causal skeleton.
	
	In neuroscience, it is often the case that there are unmeasured confounders. If we ignore the confounders, we may have spurious edges. The two-step method can be extended to cover the confounder case, by replacing equation (\ref{FCM_matrix}) with
	\begin{equation*}
	\mathbf{x}^{(i)} = B^{(i)} \mathbf{x}^{(i)} + M^{(i)} C^{(i)} + U^{(i)},
	\end{equation*}
	where $M^{(i)}$ is a vector indicating which observed variables are affected by unmeasured confounders, $C^{(i)}$ is a vector of unmeasured confounders, if they exist, and $U^{(i)}$ is the vector of mutually independent noise terms. By reorganizing the above equation, we have
	\begin{equation*}
	\mathbf{x}^{(i)} = (I-B^{(i)})^{-1} ( M^{(i)} C^{(i)} + U^{(i)}),
	\end{equation*}
	where $ M^{(i)} C^{(i)} + U^{(i)}$ defines the unmeasured components, which are not necessarily mutually independent because of the possible presence of unmeasured confounders in $C^{(i)}$. Thus, generally speaking, it does not follow the ICA model. Instead, the equation corresponds to the independent subspace analysis (ISA) model \cite{ISA1, ISA2}: The components in $ M^{(i)} C^{(i)} + U^{(i)}$ can be divided into mutually independent variables or groups of variables, where the variables within the same group are not independent.  Each of these groups is a set of confounders plus noise terms that influence one another or are influenced by some common unknown mechanism. Under mild assumptions, the solution to ISA can be found by
	applying ICA and then testing for the independence between the ICA outputs. Otherwise, if there are no confounders, the equation is a standard ICA model.

	\chapter{Autism Spectrum Disorder Diagnosis on ABIDE Dataset}
	In this chapter, we identify atypical connectivities on Autism Brain Imaging Data Exchange (ABIDE) dataset for ASD diagnosis (Chapter \ref{Ch4_Sec: Datasets}). Particularly, we use the two-step method to learn causal influence strength as candidate features and further perform feature selection and support vector machine for diagnosis (Chapter \ref{Ch4_Sec: Methods}). Experimental results demonstrate the effectiveness of the proposed methods (Chapter \ref{Ch4_Sec: Results}).
	
	\section{Datasets} \label{Ch4_Sec: Datasets}
	We used resting-state functional magnetic resonance imaging (R-fMRI) data from ABIDE \RNum{1} dataset \cite{ABIDE1}. ABIDE \RNum{1} is a publicly available dataset that involves R-fMRI, structural magnetic resonance imaging, and phenotype information for individuals with ASD and TCs, aggregated from 16 international sites. Briefly, ASD was determined by either 1) combining clinical judgment with ‘gold standard’ diagnostic instruments--ADOS and/or ADI-R or 2) using these ‘gold standard’ diagnostic instruments only. Across collections, R-fMRI acquisition durations varied from five to eight minutes per individual. In most collections, individuals were verbally asked to keep their eyes open. All R-fMRI data were acquired using 3 Tesla scanners. Sequence parameters of R-fMRI datasets at each data collection site vary and can be found in \cite{ABIDE1}. 
	
	We used a publicly available preprocessed version of this dataset provided by the Preprocessed Connectome Project initiative \cite{ABIDE_preprocess}. Specifically, we used the data processed by Connectome Computation System (CCS). 
	Data were further selected based on the data quality. For quality assessment, the Preprocessed Connectome Project quality assurance protocol was used \cite{PCP}. They include spatial metrics of scanner performance, such as incomplete brain coverage, ghosting, contrast to noise ratio, and artifactual voxel detection, and temporal metrics to quantify head motion \cite{ABIDE1}.
	
	To reduce the dimensionality of the estimated connectomes and to improve the interpretability of results, connectome nodes were defined from regions of interest (ROIs). Particularly, we applied the Automated Anatomical Labeling (AAL) atlas on the voxel data, resulting in 116 ROIs. The time series of each node for analysis is the averaged voxel signals within each ROI.
	
	For statistical reliability, we chose the datasets with the largest sample sizes. We used three datasets from University of Michigan (48 TCs and 30 individuals with ASD), from New York University (86 TCs and 74 individuals with ASD), and from University of Utah, School of Medicine (26 TCs and 43 individuals with ASD). See Appendix for subjects used in our experiments.
	
	\section{Methods} \label{Ch4_Sec: Methods}
	We used the two-step method to estimate causal influence strength. Then with the estimated causal influence strength as candidate features, we further performed feature selection and classification for ASD diagnosis. 
	
	\subsection{Two-Step Method for Causal Influence Estimation} \label{Ch4_Sec: two_step}
	We used the two-step method, which is introduced in Chapter 3, to learn the causal connectivities between ROIs for each subject. The two-step method is able to recover the whole causal structure efficiently, including feedbacks among brain regions. It contains two steps. In the first step, it learns a superset of the underlying causal skeleton. The results from the first step are used as constraints of connections in the second step, which makes the estimation faster and more reliable, especially in the high-dimensional case. In the second step, it identifies causal structure and causal influence strength between ROIs for each subject. Particularly, the second step relies on constrained functional causal model, which represents the effect as a linear function of the direct causes and an independent noise term:
	\begin{equation*}
	X = BX + U.
	\end{equation*}
	The resulting causal structure is uniquely identifiable under certain conditions. 
	
	Following the procedures and notations described in Chapter 3, we applied the two-step method to the three datasets separately. In the first step, the regularization parameter $\lambda$ in adaptive lasso is set $\lambda = 0.01$. The threshold $\alpha$ on $\hat{\beta}$ is set $\alpha = 0.01$; if $\hat{\beta}_{jk} > \alpha$, $O_{jk} = 1$, and otherwise, $O_{jk} = 0$.
	In the second step, the regularization parameters in equation (\ref{ICA_obj_updated}) is set $\lambda = 0.5 $ and $\gamma = 0.5$. The parameter $m$ in $\tanh(mw)$, which is used to approximate $|w|$, is set $m = 50$. The above hyperparameters follow a general configuration. A better way may be to use cross-validation, but it is time-consuming. Finally, the estimated causal adjacency matrix is denoted as $B$, with $B_{jk}$ indicates the direct causal influence strength from ROI $k$ to ROI $j$.
	
	We further considered indirect causal influences between two ROIs. For example, we used $B+B^2$ to indicate the sum of direct and one-step indirect causal influences, $B+B^2+B^3$ to indicate direct and within two-step indirect causal influences, and $B+B^2+B^3 + \cdots$ to indicate total influence strength containing direct and all indirect causal influences; mathematically, $B+B^2+B^3 + \cdots$ can be efficiently estimated by $(I-B)^{-1}B$.
	%We also considered the variance of noise term $U$ to indicate self-loop strength, and power spectrum of $U$ to indicate frequency information of dynamic connectivity. 
	Besides causal influences, we also considered other information in the causal model, e.g., variance of estimated noise terms $U$ and power spectrum of $U$, which indicates frequency information of dynamic connectivity. Since the power spectrum of $U$ is a sequence, we extracted the curvature of the power spectrum by fitting a second-order polynomial regression function and used the coefficient to the quadratic term as an indicator of the power spectrum.
	%Note that the proposed two-step method assumes that there are no self-loops and the data is stationary in estimating direct causal connections $B$. However, in real fMRI data, these assumptions might be violated. Therefore, after applying the two-step method, we further examined the estimated noise $U$; particularly, we used variance of $U$ to indicate self-loop strength and power spectrum of $U$ to indicate frequency information of dynamic connectivity.
	
	\subsection{Feature Selection}  \label{Ch4_Sec: feature_selection}
	In the case of high feature dimensionality and limited sample size, feature selection is an effective way to circumvent curse of dimensionality and enhance generalization by reducing overfitting. In our experiments, we performed both feature ranking and wrapper methods for feature selection.
	
	\subsubsection{Feature ranking}
	The estimated causal adjacency strength $B$ from the two-step method are used as features for diagnosis. With AAL atlas, we have 116 ROIs, resulting in $116 \times 116$ total number of features. With limited number of subjects in the fMRI recording, it may lead to be overfitting. Thus, we first preformed feature selection to remove uninformative features to avoid overfitting. 
	
	Particularly, we applied two-sample t-test to choose a subset of features which are useful for prediction. The two-sample t-test is used to compare whether the average difference between two groups is really significant or is due to random error.
	
	Let $\mu_c$ and $v_c$ represent the mean and variance of causal adjacency strength of TCs, with
	\begin{equation}
	\begin{array}{l}
	\mu_c = \frac{1}{|S_{\text{TC}}|} \sum_{i \in S_{\text{TC}}} B^{(i)}, \\
	v_c = \frac{1}{|S_{\text{TC}}|} \sum_{i \in S_{\text{TC}}} (B^{(i)}-\mu_c)^2,
	\end{array}
	\end{equation}
	where $S_{\text{TC}}$ indicates the set of subject ID in the group of TC, and $|S_{\text{TC}}|$ represents the number of TCs. Similarly, we estimate the mean $\mu_a$ and variance $v_a$ of causal adjacency strength of individuals with ASD, with
	\begin{equation}
	\begin{array}{l}
	\mu_a = \frac{1}{|S_{\text{ASD}}|} \sum_{i \in S_{\text{ASD}}} B^{(i)}, \\
	v_a = \frac{1}{|S_{\text{ASD}}|} \sum_{i \in S_{\text{ASD}}} (B^{(i)}-\mu_a)^2,
	\end{array}
	\end{equation}
	where $S_{\text{ASD}}$ indicates the set of subject ID in the group of ASD, and $|S_{\text{ASD}}|$ represents the number of individuals with ASD.
	Particularly, for the causal influence from ROI $k$ to ROI $j$, the t-statistic can be calculated as
	\begin{equation}
	t_{jk} = \frac{\mu_{t,jk} - \mu_{a,jk}}{ \sqrt{v_{jk} \cdot (\frac{1}{|S_\text{TC}|} + \frac{1}{|S_\text{ASD}|}})},
	\end{equation}
	where $v_{jk} = \frac{(|S_\text{TC}|-1){v_{c,jk}} + (|S_\text{ASD}|-1){v_{a,jk}}}{|S_\text{TC}|+|S_\text{ASD}|-2}$. We chose the subset of features whose $p$ values are smaller than $0.05$.
	
	Furthermore, we considered other sets of features, including indirect causal influences $B+B^2$ and $B+B^2+B^3$, total causal influence $B+B^2+B^3 + \cdots$, variance of noise term $U$, and frequency information of dynamic connectivity. For all those features, we performed the same two-sample t-test for feature selection and chose the subset of features with significance level $0.05$.
	
	\subsubsection{Wrapper methods}
	Feature ranking does not consider joint effects between features, where selected features may be redundant. It has been shown that selecting the most relevant features is usually suboptimal for building a predictor, particularly if the features are redundant \cite{Wrapper1,Wrapper2, Feature_selection_Guyon}. Thus, after feature ranking, we performed wrapper methods, which use the prediction performance of a given learning machine to assess the relative usefulness of subsets of features. In practice, one needs to define: (i) how to search the space of all possible feature subsets; (ii) how to assess the prediction performance of a learning machine to guide the search and halt it; and (iii) which predictor to use \cite{Feature_selection_Guyon}.
	
	Specifically, in our experiments, we exploited forward greedy search strategy to search the space of possible feature subsets; features are progressively incorporated into larger and larger subsets in a greedy way. It has been shown that greedy search strategies are computationally advantageous and are robust against overfitting \cite{Feature_selection_Guyon}. We used out-of-sample prediction performance to evaluate the usefulness of a subset of features, and the search procedure is halted when the prediction performance no longer improves. We used support vector machine to do prediction, which will be introduced in the next section.
	
	In wrapper methods, we considered different sets of features:
	\begin{itemize}[noitemsep,nolistsep] 
		\item Direct causal influences $B$, variance of noise term, and frequency information of dynamic connectivity.
		\item Direct causal influences $B$.
		\item Variance of noise term.
		\item Frequency information of dynamic connectivity.
	\end{itemize}

	\subsection{Support Vector Classification}  \label{Ch4_Sec: SVM}
	With the available phenotype information from ABIDE dataset, diagnosis between TCs and individuals with ASD becomes a supervised learning problem. Specifically, we applied support vector machine (SVM) \cite{SVM} on the selected features to do classification.
	
	Denoted by $\mathbf{c}_i$ the set of selected features and by $y_i$ its label information for the $i$th subject. Given a set of pairs $(\mathbf{c}_i, y_i)$ with $i = 1, . . . , n$ and $y_i \in \{1,-1\}$, the SVM aims to solve the following optimization problem:
	\begin{equation*}
	\begin{array}{ll}
	\min_{\mathbf{w},b,\zeta} & \frac{1}{2} \mathbf{w}^T \mathbf{w} + C \sum_{i=1}^{n} \zeta_i,\\
	\text{subject to} & y_i (\mathbf{w}^T \phi(\mathbf{c}_i)+b ) \geq 1-\zeta_i,\\
	& \zeta_i \geq 0,
	\end{array}
	\end{equation*}
	where $C>0$ is the penalty parameter of the error term, and the training vector $\mathbf{c}_i$ is mapped into a higher dimensional space by the function $\phi$.  
	With the kernel trick $k(\mathbf{c}_i,\mathbf{c}_j) = \langle \phi(\mathbf{c}_i), \phi(\mathbf{c}_j) \rangle$, we do not need to learn $\phi$ explicitly. We used radial basis functions as kernels to learn a nonlinear classification rule, where $k(\mathbf{c}_i,\mathbf{c}_j) = \exp (-\frac{|| \mathbf{c}_i-\mathbf{c}_j||}{2\sigma^2})$, and $\sigma$ denotes the kernel width.
	
	%In our experiments, the kernel width $\sigma$ is chosen by ..
	
	\subsection{Implementation and Validation}
	We first learned causal adjacency strengths between ROIs using the two-step method (Chapter \ref{Ch4_Sec: two_step})  as features, and then we did feature selection (Section \ref{Ch4_Sec: feature_selection}) to remove uninformative features to avoid overfitting. We then performed classification between TCs and individuals with ASD by SVM (Section \ref{Ch4_Sec: SVM}) using selected features. The algorithms were implemented in Matlab R2017a.
	
	To estimate the prediction accuracy, we divided the data into training set and test set and performed leave-one-out prediction for model evaluation; each subject is used as the test set in turn. It is imperative to avoid using the label information from test subjects for training the model as this would result positively biased estimates of the prediction accuracy. 
	Suppose that we have a set of pairs $\{(X^{(i)}, y_i)\}_{i=1}^n$, where $X^{(i)}$ is the BOLD signal, and $y_i \in \{1,-1\}$ indicates the label; $y_i=-1$ means that the corresponding subject is autistic, and $y_i=1$ means that the corresponding subject is typical control. Let $S$ represent the IDs of all subjects with $|S|=n$, $S_{\text{train}}$ the IDs of training set,  and $S_{\text{test}}$ the IDs of test set. Algorithm \ref{Algo: implementation} gives the diagnostic procedure of feature learning, feature selection, and classification.
	
	\begin{algorithm}[htp!] % enter the algorithm environment
		\caption{Diagnostic procedure}
		\begin{algorithmic}
			\State	\textbf{INPUT:} $\{(X^{(i)}, y_i)\}_1^n$
			\State \textbf{Output:} leave-one-out prediction accuracy: Acc
			
			\item []
			\item \%\% Feature learning with two-step methods
			\State $\{B^{(i)}\}_{i=1}^n =$ two-step $(\{X^{(i)}\}_1^n)$

			\item[]
			\item \%\% Feature selection and classification
			\State Initialization: $\text{Acc}=0$
			\For{$i=1$ to $n$}  
			\State $S_\text{test} = i$   %\qquad \qquad \% ids of test subject
			\State $S_\text{train} = S/S_\text{test}$ %  \qquad \qquad \% ids of training subject
			\State Denote $(B_\text{test}, y_\text{test})$ as the feature of subjects $S_\text{test}$
			\State Denote $(B_\text{train}, y_\text{train})$ as the feature of subjects $S_\text{train}$
			\item[]
			\State Subset of features $\mathbf{c}= \text{feature ranking}(B_\text{train}, y_\text{train})$ 
			\State Subset of features $\mathbf{c}'= \text{wrapper method}(\mathbf{c}, y_\text{train})$ 
			\item[]
			\State $\text{model} = \text{SVM}(\mathbf{c}', y_\text{train} )$
			\State $\hat{y}_\text{test} = \text{model}.\text{predict}(X_\text{test})$
			\State $\text{Acc} = \text{Acc} + (\hat{y}_\text{test} == y_\text{test})$
			\EndFor
			\State $\text{Acc} = \text{Acc}/n$
		\end{algorithmic}
		\label{Algo: implementation}
	\end{algorithm} % enter the algorithm environment

	\section{Experimental Results} \label{Ch4_Sec: Results}
	In this section, we present the most notable trends emerging from our analyses, including intra-site and inter-site diagnostic accuracy, significant features for diagnosis, and properties of brain connectivity structure. 
	
	\subsection{Intra-Site Diagnostic Accuracy}
	To evaluate the intra-site diagnostic accuracy, we separated each dataset into training set and test set, by training the model exclusively on training set and predicting on the test set; particularly, we used leave-one-out prediction performance for accuracy estimation. We used the direct causal connections $B$, the variance of noise term, and the frequency information of dynamic networks as features. To avoid overfitting, we reduced the number of features by both feature ranking (two-sample t-test) and wrapper methods (forward feature selection) in Section \ref{Ch4_Sec: feature_selection}. Then the classification was performed by SVM introduced in Section \ref{Ch4_Sec: SVM}. Table \ref{Table: Intra-site}(a) gives the intra-site out-of-sample prediction accuracy on the three datasets.  Besides the overall accuracy, it also gives the specificity and sensitivity, which measure the proportion of TCs and the proportion of ASD that are correctly identified, respectively. We compared with using Pearson's correlation and partial correlation as features for classification in Table \ref{Table: Intra-site}(b) and Table \ref{Table: Intra-site}(c), respectively. For the partial correlation, we used the implementation from \cite{L1precision}. When evaluating the performance of correlation and partial correlation-based classification, we performed the same procedures as our methods in feature selection and classification. From experimental results shown in Table \ref{Table: Intra-site}, with estimated causal model as features, it achieves much better diagnostic accuracy than using correlation and partial correlation on all datasets, improving more than 10\%.

	\begin{table}[htp!]
		\centering
		\begin{tabular}{llll}
			\multicolumn{4}{c}{(a) Causality}                                                                                                                                                                                 \\ \hline
			\multicolumn{1}{|l|}{}    & \multicolumn{1}{c|}{Accuracy} & \multicolumn{1}{l|}{\begin{tabular}[c]{@{}l@{}}Specificity\\ (TC)\end{tabular}} & \multicolumn{1}{l|}{\begin{tabular}[c]{@{}l@{}}Sensitivity\\ (ASD)\end{tabular}} \\ \hline
			\multicolumn{1}{|l|}{NYU} & \multicolumn{1}{l|}{82\%}     & \multicolumn{1}{l|}{87\%}                                                       & \multicolumn{1}{l|}{76\%}                                                        \\ \hline
			\multicolumn{1}{|l|}{UM}  & \multicolumn{1}{l|}{87\%}     & \multicolumn{1}{l|}{92\%}                                                       & \multicolumn{1}{l|}{80\%}                                                        \\ \hline
			\multicolumn{1}{|l|}{USM} & \multicolumn{1}{l|}{83\%}     & \multicolumn{1}{l|}{69\%}                                                       & \multicolumn{1}{l|}{91\%}                                                        \\ \hline
			&                               &                                                                                 &                                                                                  \\
			\multicolumn{4}{c}{(b) Correlation}                                                                                                                                                                               \\ \hline
			\multicolumn{1}{|l|}{}    & \multicolumn{1}{l|}{Accuracy} & \multicolumn{1}{l|}{\begin{tabular}[c]{@{}l@{}}Specificity\\ (TC)\end{tabular}} & \multicolumn{1}{l|}{\begin{tabular}[c]{@{}l@{}}Sensitivity\\ (ASD)\end{tabular}} \\ \hline
			\multicolumn{1}{|l|}{NYU} & \multicolumn{1}{l|}{73\%}     & \multicolumn{1}{l|}{74\%}                                                       & \multicolumn{1}{l|}{70\%}                                                        \\ \hline
			\multicolumn{1}{|l|}{UM}  & \multicolumn{1}{l|}{72\%}     & \multicolumn{1}{l|}{83\%}                                                       & \multicolumn{1}{l|}{53\%}                                                        \\ \hline
			\multicolumn{1}{|l|}{USM} & \multicolumn{1}{l|}{68\%}     & \multicolumn{1}{l|}{54\%}                                                       & \multicolumn{1}{l|}{77\%}                                                        \\ \hline
			&                               &                                                                                 &                                                                                  \\
			\multicolumn{4}{c}{(c) Partial correlation}                                                                                                                                                                       \\ \hline
			\multicolumn{1}{|l|}{}    & \multicolumn{1}{l|}{Accuracy} & \multicolumn{1}{l|}{\begin{tabular}[c]{@{}l@{}}Specificity\\ (TC)\end{tabular}} & \multicolumn{1}{l|}{\begin{tabular}[c]{@{}l@{}}Sensitivity\\ (ASD)\end{tabular}} \\ \hline
			\multicolumn{1}{|l|}{NYU} & \multicolumn{1}{l|}{71\%}         & \multicolumn{1}{l|}{72\%}                                                           & \multicolumn{1}{l|}{69\%}                                                            \\ \hline
			\multicolumn{1}{|l|}{UM}  & \multicolumn{1}{l|}{62\%}         & \multicolumn{1}{l|}{77\%}                                                           & \multicolumn{1}{l|}{37\%}                                                            \\ \hline
			\multicolumn{1}{|l|}{USM} & \multicolumn{1}{l|}{64\%}         & \multicolumn{1}{l|}{20\%}                                                           & \multicolumn{1}{l|}{91\%}                                                            \\ \hline
		\end{tabular}
		\caption{Intra-site prediction accuracy on the three datasets, with (a) estimated causal model, (b) correlation, and (c) partial correlation as features for classification.}
		\label{Table: Intra-site}
	\end{table}

	We also compared the diagnostic accuracy with different sets of features. Particularly, we considered the set of features with only direct causal influence strength $B$, the set with only noise variance, and the set with only frequency information in Table \ref{Table: Intra-diffsets}. It can be seen that with only direct causal influence strength as features, it does not make much difference in diagnostic accuracy; on UM and USM dataset it remains the same, while on NYU dataset it reduces a bit, from 82\% to 79\%. However, with only variance of noise term and only frequency information, it is less accurate. Therefore, the variance of noise term and frequency information may be less informative in diagnosis by themselves, compared to causal connectivities.

	\begin{table}[htp!]
		\centering
		\begin{tabular}{llll}
			\multicolumn{4}{c}{(a) Causal adjacency strength}                                                                                                                                                                        \\ \hline
			\multicolumn{1}{|l|}{}    & \multicolumn{1}{c|}{Accuracy} & \multicolumn{1}{l|}{\begin{tabular}[c]{@{}l@{}}Specificity\\ (TC)\end{tabular}} & \multicolumn{1}{l|}{\begin{tabular}[c]{@{}l@{}}Sensitivity\\ (ASD)\end{tabular}} \\ \hline
			\multicolumn{1}{|l|}{NYU} & \multicolumn{1}{l|}{79\%}     & \multicolumn{1}{l|}{83\%}                                                       & \multicolumn{1}{l|}{76\%}                                                        \\ \hline
			\multicolumn{1}{|l|}{UM}  & \multicolumn{1}{l|}{87\%}     & \multicolumn{1}{l|}{92\%}                                                       & \multicolumn{1}{l|}{80\%}                                                        \\ \hline
			\multicolumn{1}{|l|}{USM} & \multicolumn{1}{l|}{83\%}     & \multicolumn{1}{l|}{69\%}                                                       & \multicolumn{1}{l|}{91\%}                                                        \\ \hline
			&                               &                                                                                 &                                                                                  \\
			\multicolumn{4}{c}{(b) Variance of noise term}                                                                                                                                                                                     \\ \hline
			\multicolumn{1}{|l|}{}    & \multicolumn{1}{l|}{Accuracy} & \multicolumn{1}{l|}{\begin{tabular}[c]{@{}l@{}}Specificity\\ (TC)\end{tabular}} & \multicolumn{1}{l|}{\begin{tabular}[c]{@{}l@{}}Sensitivity\\ (ASD)\end{tabular}} \\ \hline
			\multicolumn{1}{|l|}{NYU} & \multicolumn{1}{l|}{64\%}     & \multicolumn{1}{l|}{70\%}                                                       & \multicolumn{1}{l|}{58\%}                                                        \\ \hline
			\multicolumn{1}{|l|}{UM}  & \multicolumn{1}{l|}{71\%}     & \multicolumn{1}{l|}{88\%}                                                       & \multicolumn{1}{l|}{43\%}                                                        \\ \hline
			\multicolumn{1}{|l|}{USM} & \multicolumn{1}{l|}{58\%}     & \multicolumn{1}{l|}{27\%}                                                       & \multicolumn{1}{l|}{78\%}                                                        \\ \hline
			&                               &                                                                                 &                                                                                  \\
			\multicolumn{4}{c}{(c) Frequency information of dynamic connectivity}                                                                                                                                                          \\ \hline
			\multicolumn{1}{|l|}{}    & \multicolumn{1}{l|}{Accuracy} & \multicolumn{1}{l|}{\begin{tabular}[c]{@{}l@{}}Specificity\\ (TC)\end{tabular}} & \multicolumn{1}{l|}{\begin{tabular}[c]{@{}l@{}}Sensitivity\\ (ASD)\end{tabular}} \\ \hline
			\multicolumn{1}{|l|}{NYU} & \multicolumn{1}{l|}{67\%}     & \multicolumn{1}{l|}{71\%}                                                       & \multicolumn{1}{l|}{62\%}                                                        \\ \hline
			\multicolumn{1}{|l|}{UM}  & \multicolumn{1}{l|}{71\%}     & \multicolumn{1}{l|}{85\%}                                                       & \multicolumn{1}{l|}{47\%}                                                        \\ \hline
			\multicolumn{1}{|l|}{USM} & \multicolumn{1}{l|}{70\%}     & \multicolumn{1}{l|}{46\%}                                                       & \multicolumn{1}{l|}{84\%}                                                        \\ \hline
		\end{tabular}
		\caption{Intra-site diagnostic accuracy with different sets of features: (a) causal adjacency strength, (b) variance of noise term, and (c) frequency information of dynamic connectivity.}
		\label{Table: Intra-diffsets}
	\end{table}
	
	\subsection{Inter-Site Diagnostic Accuracy}
	
	To further evaluate the generalization performance of our methods, we evaluated the inter-site prediction accuracy, with direct causal influences $B$, the variance of noise term, and the frequency information as features. Table \ref{Table: inter-site} shows the inter-site prediction accuracy, i.e., using classifiers trained on the dataset from one site to predict the dataset on another site. For example, NYU $\rightarrow$ UM means that using classifiers trained on NYU to diagnose subjects from UM. Not surprisingly, the inter-site prediction is less accurate compared to intra-site prediction. Among intra-site predictions between different datasets, the prediction accuracies between USM and NYU are the highest. The reduction in inter-site accuracy may be due to heterogeneity between different datasets.
	We compared the prediction accuracy with baseline naive classifiers. The naive classifier is based on Zero-Rule, which simply predicts the majority class in the dataset. With naive classifier, the prediction accuracy on NYU, UM, and USM are 54\%, 62\%, and 62\%, respectively. Therefore, the classifier we trained on the learned features are informative, outperforming the naive classifier.

	\begin{table}[htp!]
		\centering
		\begin{tabular}{|l|l|l|l|}
			\hline
			& Accuracy & \begin{tabular}[c]{@{}l@{}}Specificity \\ (TCs)\end{tabular} & \begin{tabular}[c]{@{}l@{}}Sensitivity \\ (ASD)\end{tabular} \\ \hline
			USM$\rightarrow$NYU & 71\%     & 70\%                                                         & 72\%                                                         \\ \hline
			UM$\rightarrow$NYU  & 65\%     & 83\%                                                         & 45\%                                                         \\ \hline
			USM$\rightarrow$UM  & 62\%     & 63\%                                                         & 60\%                                                         \\ \hline
			NYU$\rightarrow$UM  & 67\%     & 69\%                                                         & 63\%                                                         \\ \hline
			UM$\rightarrow$USM  & 70\%     & 58\%                                                         & 77\%                                                         \\ \hline
			NYU$\rightarrow$USM & 74\%     & 77\%                                                         & 72\%                                                         \\ \hline
		\end{tabular}
		\caption{Inter-site diagnostic accuracy.}
		\label{Table: inter-site}
	\end{table}

	\subsection{Identified Significant Features}
	The classification results shown above demonstrate the effectiveness of using the features learned with our methods for diagnosis; that is, the learned features in the two groups, the group of TCs and that of ASD, are informative for diagnosis. Next, we want to know which particular features, e.g., which causal influences, are significantly different between the two groups. To this end, we did significant test on each feature separately. Specifically, we used the two-sample t-test introduced in Section \ref{Ch4_Sec: feature_selection}, with significance level 0.05.
	%Since we have already concluded that the learned features differ in the two groups  because they are informative for diagnosis, we do not need to do multiple-test correction; we make a conclusion about the significance of each feature separately. Specifically, we used the two-sample t-test introduced in Section \ref{Ch4_Sec: feature_selection}, with significance level 0.05.
	
\begin{figure}[htp!]
		\centering
		\includegraphics[width=.9\textwidth]{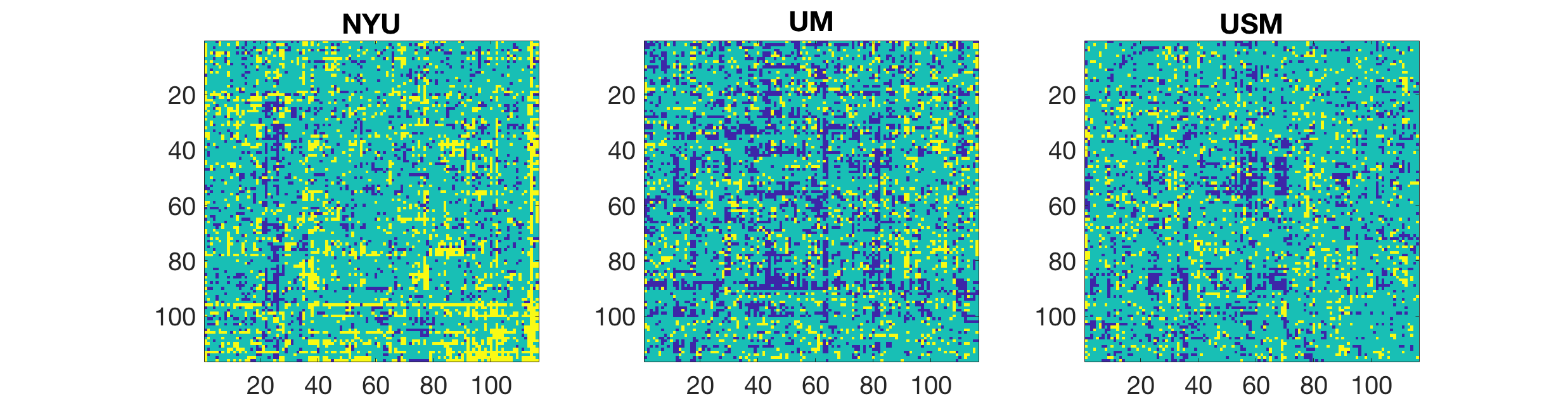}
		\caption{Identified significant direct and indirect brain influences in the three datasets: NYU dataset, UM dataset, and USM dataset. For the $i$th row and $j$th column, a yellow entry means that the direct/indirect causal influence from ROI $j$ to ROI $i$ is significantly enhanced in the group of ASD; blue means that the causal influence is significantly reduced; green means that the causal influence is not significantly different.}
		\label{Fig: features}
	\end{figure}

	Figure \ref{Fig: features} shows the significant brain influences identified by two-sample t-test from the three datatsets. The brain influences consist of direct causal influences $B$, sum of direct and finite-step indirect causal influences $B+B^2$ and $B+B^2+B^3$, and total influence strength $B+B^2+B^3+\cdots$. For the $i$th row and $j$th column, a yellow entry indicates that the direct/indirect causal influence from ROI $j$ to ROI $i$ is significantly enhanced in the group of ASD; blue indicates that the causal influence is significantly reduced; green indicates that the causal influence is not significantly different. Note that here we did not separate direct from indirect causal influences, and there are no conflicts between them. For the NYU dataset, there are more enhanced influences than reduced ones. Specifically, the enhanced ones are mostly concentrated in the cerebellum area, and the reduced ones mostly start from the frontal lobe. For the UM dataset and USM dataset, there are more reduced influences than enhanced ones. We found that the identified significant features vary across datasets. It may be for the following reasons. 
	\begin{itemize}[noitemsep,nolistsep] 
		\item The three datasets are from different sites, with different scanners and different protocols. 
		\item The number of subjects in each dataset is relatively small, in which case the statistical reliability is largely affected if measurement noises are large.
		\item The mechanism of ASD might be much more complicated, beyond the scope of the information contained in the fMRI signal. 
	\end{itemize}
	
\begin{figure}[htp!]
		\centering
		\includegraphics[width=.45\textwidth]{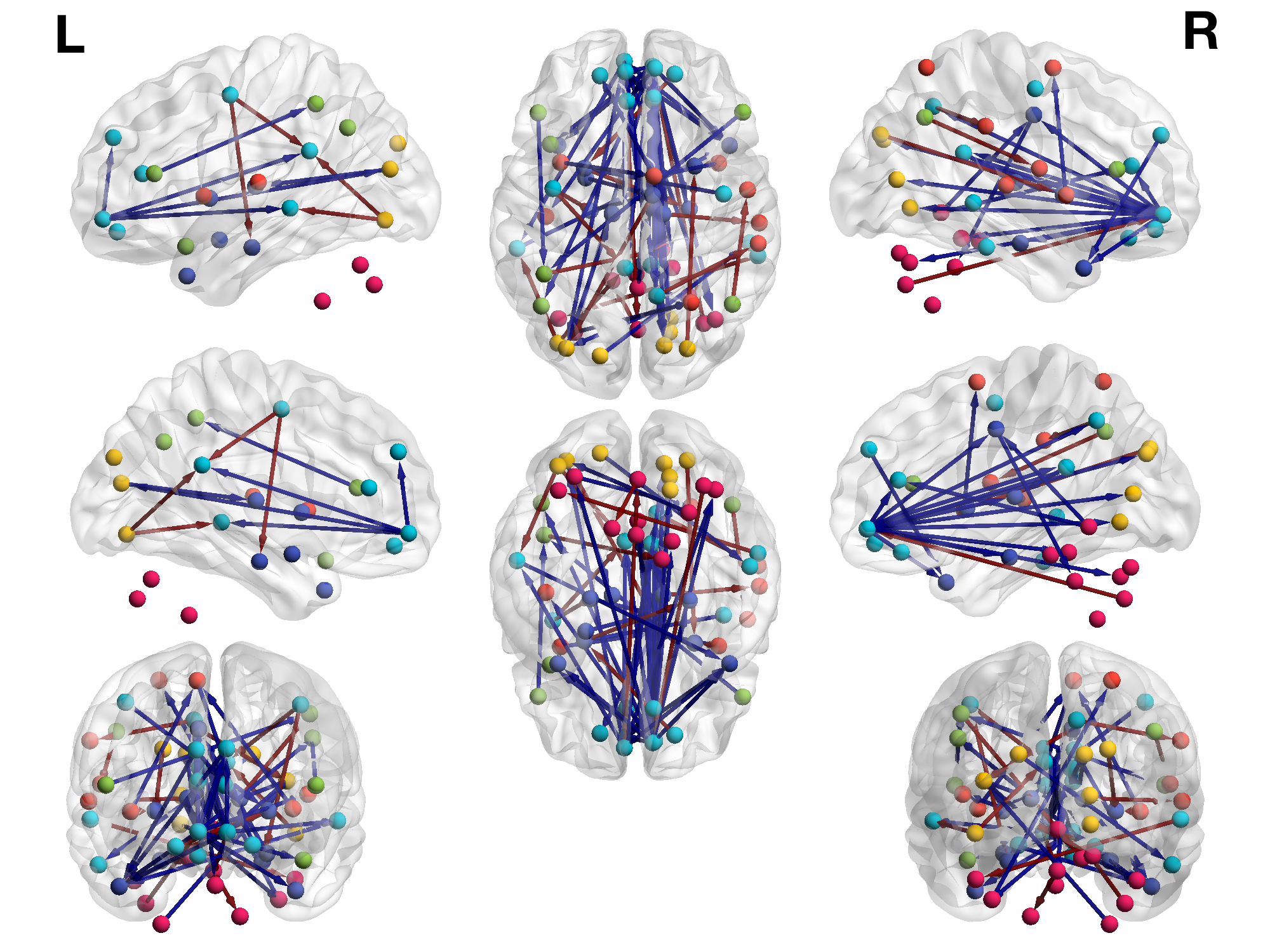}
		\caption{Identified consistent enhanced and reduced influences, from $B$, $B+B^2$, $B+B^2+B^3$, total causal influence strength, variance of noise term, frequency of dynamic connectivity. The red edge indicates enhanced influence, and the blue edge indicates reduced influence.}
		\label{Fig: consistent_features}
	\end{figure}

	Figure \ref{Fig: consistent_features} shows the identified significant features, from $B$, $B+B^2$, $B+B^2+B^3$, $B+B^2+B^3+\cdots$, variance of noise term, and frequency information, which are consistent across the three datasets. The red edge indicates enhanced influence, and the blue edge indicates reduced influence. Table \ref{Table: feature_list} gives a full list of consistently significant features. We found that most of the reduced influences are going from the frontal lobe. 
	
	\begin{table}[htp!]
		\centering
		\resizebox{0.65\columnwidth}{!}{
			\begin{tabular}{|l|}
				\hline
				\textbf{Enhanced influences     }                                                                                                                                                                                                                                                                                                                                                                                                                                                                                                                                                                                                                                                                                                                                                                                                                                                                                                                                                                                                                                                                                                                                                                                                                                                       \\ \hline
				\begin{tabular}[c]{@{}l@{}}
					Right superior frontal gyrus, orbital part $\rightarrow$ Left putamen\\ Left superior frontal gyrus, medial orbital part $\rightarrow$ Lobule VII of vermis\\ Right insula $\rightarrow$ Right putamen\\ Right middle cingulate $\rightarrow$ Left middle occipital\\ Left inferior parietal lobule $\rightarrow$ Right Lobule III of cerebellar hemisphere\\ Right precuneus $\rightarrow$ Right supramarginal gyrus\\ Left putamen $\rightarrow$ Right parahippocampal gyrus\\ Right thalamus $\rightarrow$ Right superior temporal gyrus\\ Right middle temporal gyrus $\rightarrow$ Left crus I of cerebellar hemisphere\\ Right superior occipital $\rightarrow$ Right putamen\\ Left inferior occipital $\rightarrow$ Left posterior cingulate gyrus\\ Left inferior occipital $\rightarrow$ Left middle temporal gyrus\\ Right angular gyrus $\rightarrow$ Right rolandic operculum\\ Left precentral gyrus $\rightarrow$ Left posterior cingulate gyrus\\ Left precentral gyrus $\rightarrow$ Left parahippocampal gyrus\\ Right crus II of cerebellar hemisphere $\rightarrow$ Right superior frontal gyrus, medial orbital part\\ Lobule VII of vermis $\rightarrow$ Left lobule IX of cerebellar hemisphere
				\end{tabular} \\ \hline
				\\ \hline
				\textbf{Reduced Connections }                                                                                                                                                                                                                                                                                                                                                                                                                                                                                                                                                                                                                                                                                                                                                                                                                                                                                                                                                                                                                                                                                                                                                                                                                                                            \\ \hline
				\begin{tabular}[c]{@{}l@{}}
					Right superior frontal gyrus, medial part $\rightarrow$ Left anterior cingulate gyrus \\
					Right superior frontal gyrus, medial part $\rightarrow$ Right middle temporal pole\\
					Right superior frontal gyrus, medial orbital part $\rightarrow$ Right precuneus\\
					Right superior frontal gyrus, medial orbital part $\rightarrow$ Right middle temporal pole\\
					Right superior frontal gyrus, medial orbital part $\rightarrow$ Right middle cingulate\\
					Right superior frontal gyrus, medial orbital part $\rightarrow$ Right cuneus\\
					Right superior frontal gyrus, medial orbital part $\rightarrow$ Left superior frontal gyrus, medial part \\
					Right superior frontal gyrus, medial orbital part $\rightarrow$ Left posterior cingulate gyrus\\
					Right superior frontal gyrus, medial orbital part $\rightarrow$ Right posterior cingulate gyrus\\
					Right superior frontal gyrus, medial orbital part $\rightarrow$ Left transverse temporal gyri\\
					Right superior frontal gyrus, medial orbital part $\rightarrow$ Right anterior cingulate gyrus \\
					Right superior frontal gyrus, medial orbital part $\rightarrow$ Right parahippocampal gyrus \\
					Right superior frontal gyrus, medial orbital part $\rightarrow$ Right calcarine sulcus\\
					Right superior frontal gyrus, medial orbital part $\rightarrow$ Right lingual gyrus\\
					Right superior frontal gyrus, medial orbital part $\rightarrow$ Left angular gyrus\\
					Right superior frontal gyrus, medial orbital part $\rightarrow$ Right thalamus \\
					Right superior frontal gyrus, medial orbital part $\rightarrow$ Left superior temporal pole\\
					Right superior frontal gyrus, medial orbital part $\rightarrow$ Left middle temporal pole\\
					Right superior frontal gyrus, medial orbital part $\rightarrow$ Right crus I of cerebellar hemisphere\\
					Right superior frontal gyrus, medial orbital part $\rightarrow$ Left crus II of cerebellar hemisphere\\
					Right superior frontal gyrus, medial orbital part $\rightarrow$ Right lobule IV, V of cerebellar hemisphere\\
					~\\
					Left superior frontal gyrus, medial part $\rightarrow$ Right middle temporal pole\\
					Left superior frontal gyrus, medial orbital part $\rightarrow$ Right middle temporal pole\\
					Left superior frontal gyrus, medial orbital part $\rightarrow$ Right cuneus\\
					Left superior frontal gyrus, medial orbital part $\rightarrow$ Left middle occipital\\
					Left superior frontal gyrus, medial orbital part $\rightarrow$ Right precuneus\\
					Left superior frontal gyrus, medial orbital part $\rightarrow$ Left superior frontal gyrus, medial part \\
					Left superior frontal gyrus, medial orbital part $\rightarrow$ Left posterior cingulate gyrus\\
					Left superior frontal gyrus, medial orbital part $\rightarrow$ Right posterior cingulate gyrus\\
					Left superior frontal gyrus, medial orbital part $\rightarrow$ Left middle temporal gyrus\\
					Left superior frontal gyrus, medial orbital part $\rightarrow$ Right crus I of cerebellar hemisphere\\
					~\\
					Right precentral gyrus $\rightarrow$ Left putamen\\
					Left insula $\rightarrow$ Right supplementary motor area\\
					Left superior occipital $\rightarrow$ Right inferior temporal gyrus\\
					Left middle occipital $\rightarrow$ Left thalamus\\
					Left supramarginal gyrus $\rightarrow$ Left supramarginal gyrus\\
					Left crus II of cerebellar hemisphere $\rightarrow$ Right superior parietal lobule\\
					Left area triangularis $\rightarrow$ Left inferior parietal lobule\\
					Right area triangularis $\rightarrow$ Left inferior parietal lobule\\
					Right supplementary motor area $\rightarrow$ Left superior frontal gyrus, orbital part\\
					Right insula $\rightarrow$ Right supplementary motor area\\
					Right middle cingulate $\rightarrow$ Lobule X of vermis (nodulus)\\
					Right posterior cingulate gyrus $\rightarrow$ Left anterior cingulate gyrus\\
					Right posterior cingulate gyrus $\rightarrow$ Right thalamus\\
					Left parahippocampal gyrus $\rightarrow$ Right middle cingulate\\
					Left amygdala $\rightarrow$ Right middle cingulate\\
					Left superior occipital $\rightarrow$ Right inferior temporal gyrus\\
					Left middle temporal gyrus $\rightarrow$ Right middle temporal pole\\
					Right lobule VIII of cerebellar hemisphere $\rightarrow$ Left middle occipital\\
					Lobule IV, V of vermis $\rightarrow$ Right middle cingulate	\\
					Right gyrus rectus $\rightarrow$ Left superior frontal gyrus, medial part \\
					Right posterior cingulate gyrus $\rightarrow$ Right superior frontal gyrus, medial orbital part \\
					Lobule IV, V of vermis $\rightarrow$ Right anterior cingulate gyrus\\
				\end{tabular} \\ \hline
			\end{tabular}
		}
		\caption{Lists of consistently significant brain influences.}
		\label{Table: feature_list}
	\end{table}
	
\begin{figure}[htp!]
		\centering
		\includegraphics[width=.55\textwidth]{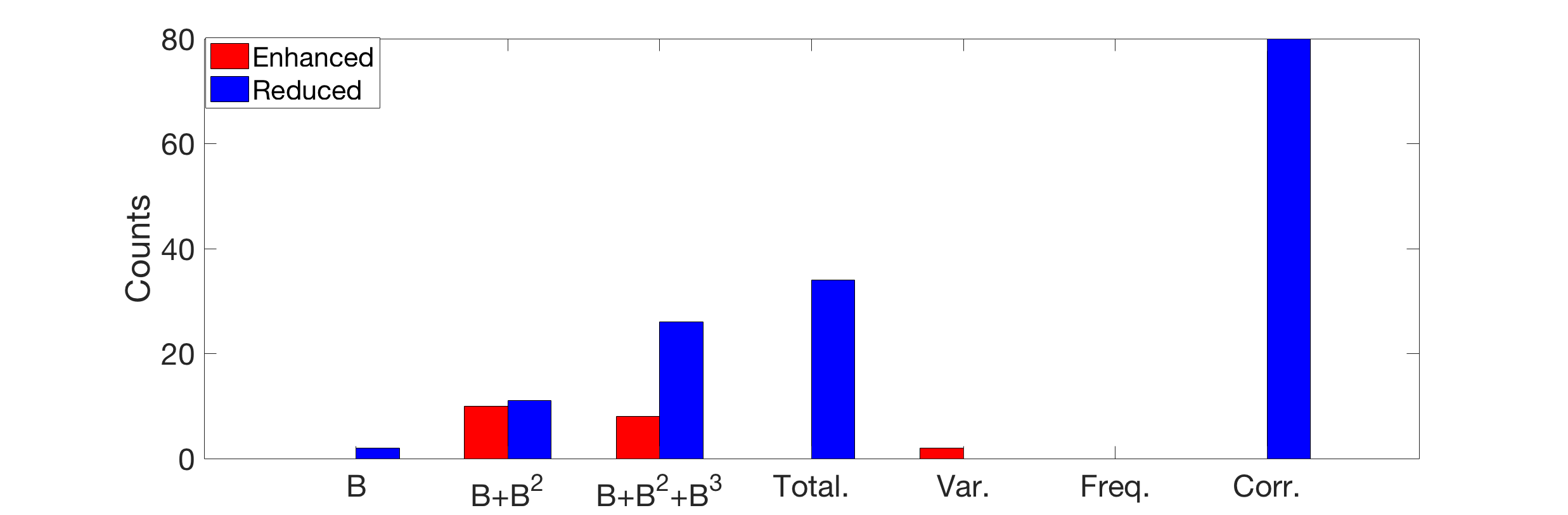}
		\caption{Number of consistently enhanced and reduced influences, from $B$, $B+B^2$, $B+B^2+B^3$, total causal influence strength, variance of noise term, frequency of dynamic connectivity, and correlation.}
		\label{Fig: num_consistent}
	\end{figure}

	Figure \ref{Fig: num_consistent} shows the number of significant features which are consistent across the three datasets; specifically, we considered enhanced and reduced influences separately. We also compared with correlation. From left to right, it shows the number of enhanced and reduced influences, from $B$, $B+B^2$, $B+B^2+B^3$, total causal influence strength, variance of noise term, frequency of dynamic connectivity, and correlation, in sequence. We found that there are more consistently reduced influences than enhanced ones. In the sets of features learned from our methods, the total causal strength has the largest number of consistent features, but only with reduced influences and no enhanced ones. %Interestingly, we found that total causal strength and correlation share the property that both of them only have consistently reduced connections; the reason might be that the total causal strength is close to correlation in some cases.
	
	Table \ref{Table: consistent-classifiction} shows the out-of-sample prediction accuracy only with consistent diagnostic features. The accuracy is evaluated by leave-one-out prediction. With only consistent features, the diagnostic accuracy is reduced on all three datasets. It also suggests that the heterogeneity across datasets may affect the results.
	\begin{table}[htp!]
		\centering
		\begin{tabular}{|l|l|l|l|}
			\hline
			& Accuracy & \begin{tabular}[c]{@{}l@{}}Specificity \\ (TCs)\end{tabular} & \begin{tabular}[c]{@{}l@{}}Sensitivity \\ (ASD)\end{tabular} \\ \hline
			NYU & 70\%     & 81\%                                                         & 68\%                                                         \\ \hline
			UM  & 69\%     & 71\%                                                         & 64\%                                                         \\ \hline
			USM & 70\%     & 63\%                                                         & 72\%                                                         \\ \hline
		\end{tabular}
		\caption{Accuracy with only consistent features.}
		\label{Table: consistent-classifiction}
	\end{table}

	\subsubsection{Reduced influences from frontal lobe}	
\begin{figure}[htp!]
		\centering
		\includegraphics[width=.5\textwidth]{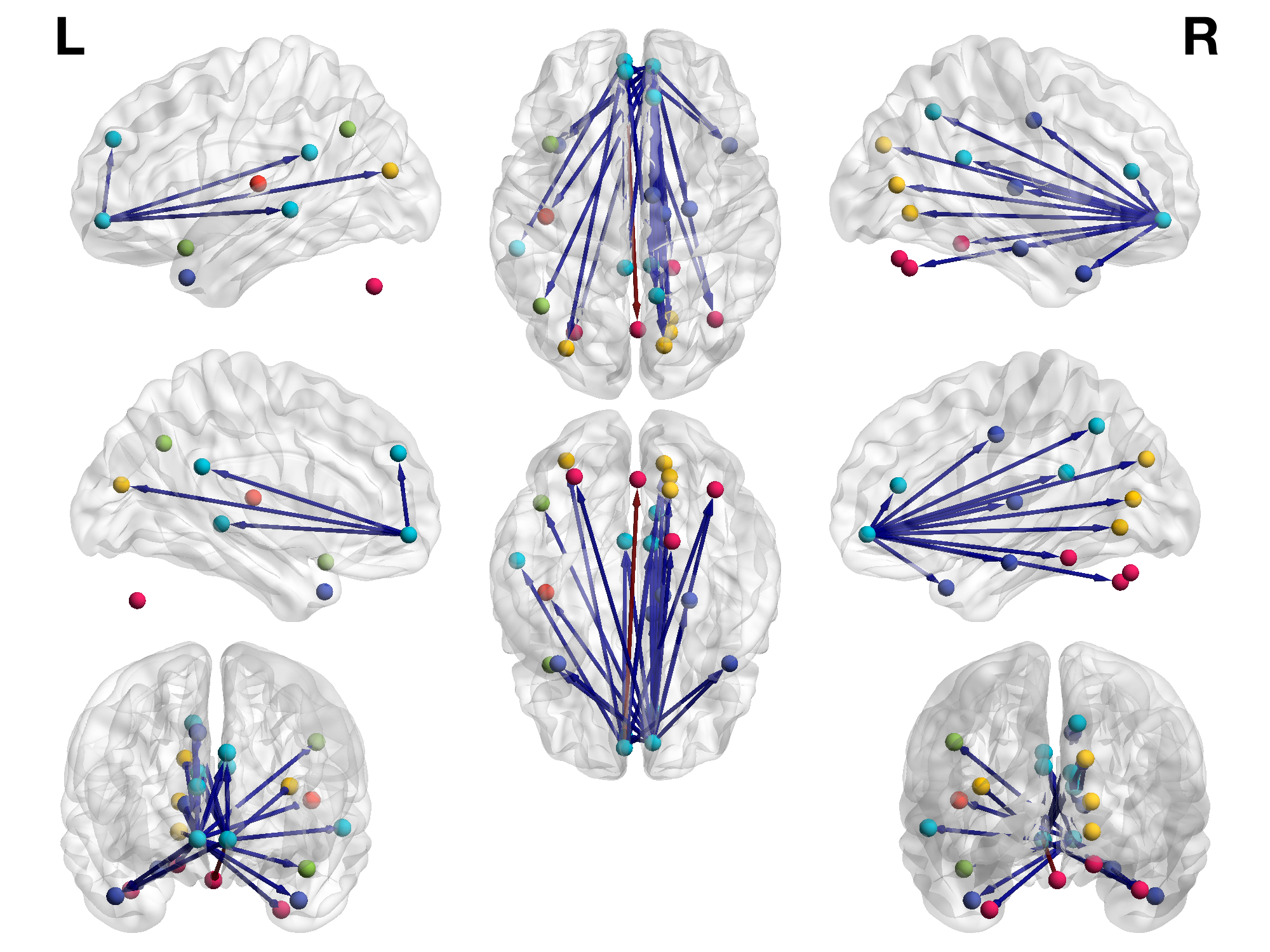}
		\caption{Identified consistent features starting from the superior frontal gyrus (medial orbital part). The features include $B$, $B+B^2$, $B+B^2+B^3$, total causal influence strength. The red edge indicates enhanced influence, and the blue edge indicates reduced influence.}
		\label{Fig: consistent_features_frontMedial}
	\end{figure}

	We found that influences starting from the frontal lobe to other areas are reduced in the group of ASD on all three datasets, especially for those influences from the medial orbital part of superior frontal gyrus. %Figure [] and Figure [] shows the t map for the connection from left superior frontal gyrus (medial orbital part) and right superior frontal gyrus (medial orbital part), respectively, to other ROIs; for a  better illustration, we map the connections to each ROI to brain surface. ...(DO IT LATER!!)
	Figure \ref{Fig: consistent_features_frontMedial} shows the significant brain influences from the superior frontal gyrus (medial orbital part) to other ROIs. For most of the influences starting from the superior frontal gyrus (medial orbital part), the causal strength is reduced. Specifically, a large set of target ROIs are concentrated in the posterior part, e.g.,the  ingual gyrus, temporal lobe, precuneus, and cerebellum. Moreover, the influences to some areas of the default mode network are reduced, such as the posterior cingulate gyrus, precuneus, angular gyrus, and parahippocampal gyrus. 

	\subsubsection{Enhanced influences from posterior to anterior or locally}	
\begin{figure}[htp!]
		\centering
		\includegraphics[width=.55\textwidth]{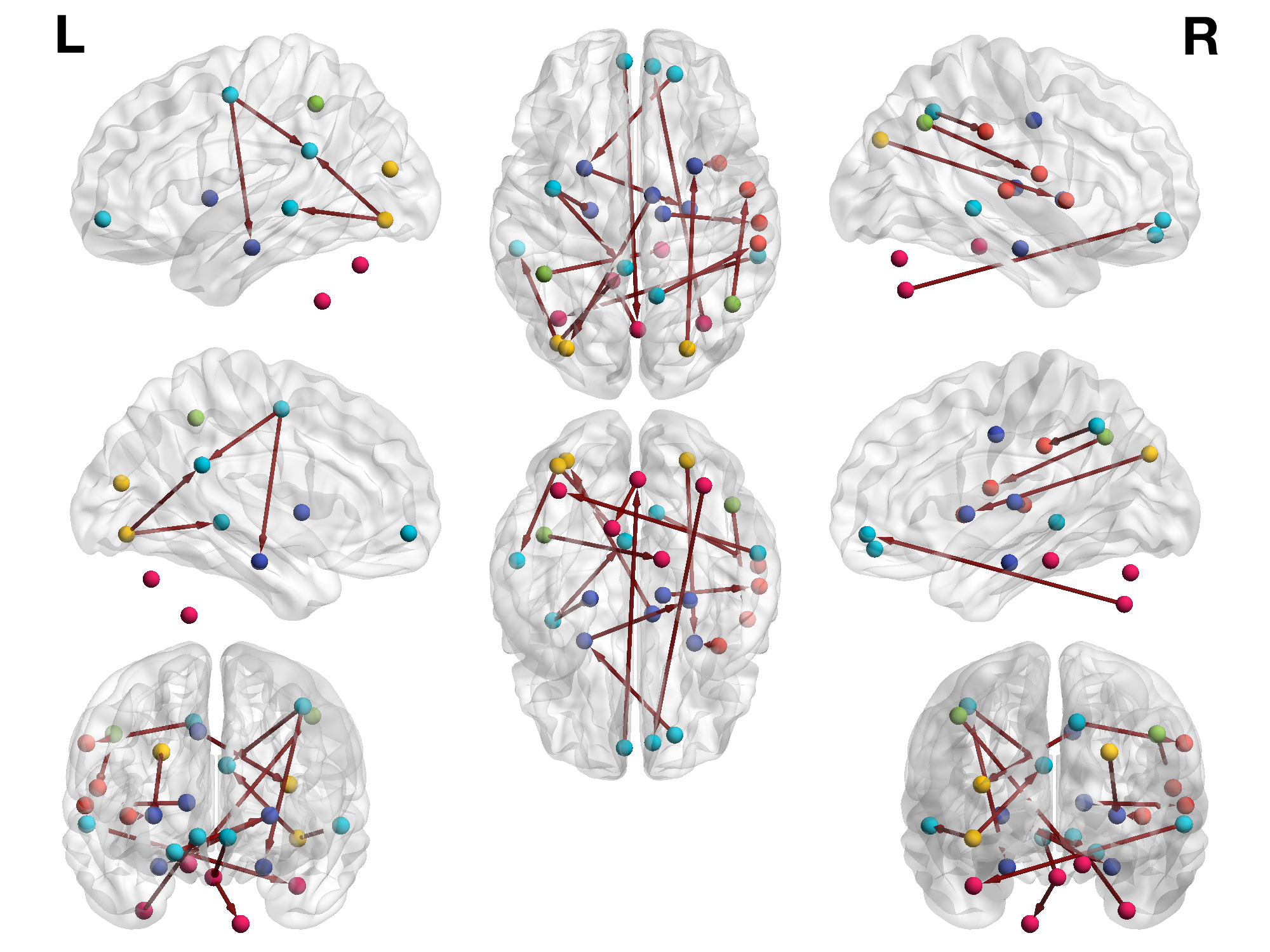}
		\caption{Consistently enhanced influences.}
		\label{Fig: share_enhance}
	\end{figure}

	We found that most enhanced influences are from posterior to anterior or in local areas. Examples of enhanced influences from posterior to anterior include Right crus II of cerebellar hemisphere $\rightarrow$ Right superior frontal gyrus and Right superior occipital $\rightarrow$ Right putamen. Examples of enhanced local influences include Lobule VII of vermis $\rightarrow$ Left lobule IX of cerebellar hemisphere, Right insula $\rightarrow$ Right putamen, and Left putamen $\rightarrow$ Right parahippocampal gyrus.
	Figure \ref{Fig: share_enhance} show all enhanced influences in brain views.
	
	Furthermore, some local influences are reduced. We noted that all identified local influences to the supplementary motor area, parietal lobule, and thalamus are reduced.
	
	\subsubsection{Reduced long rang influences and enhanced local influences}
\begin{figure}[htp!]
		\centering
		\includegraphics[width=.7\textwidth]{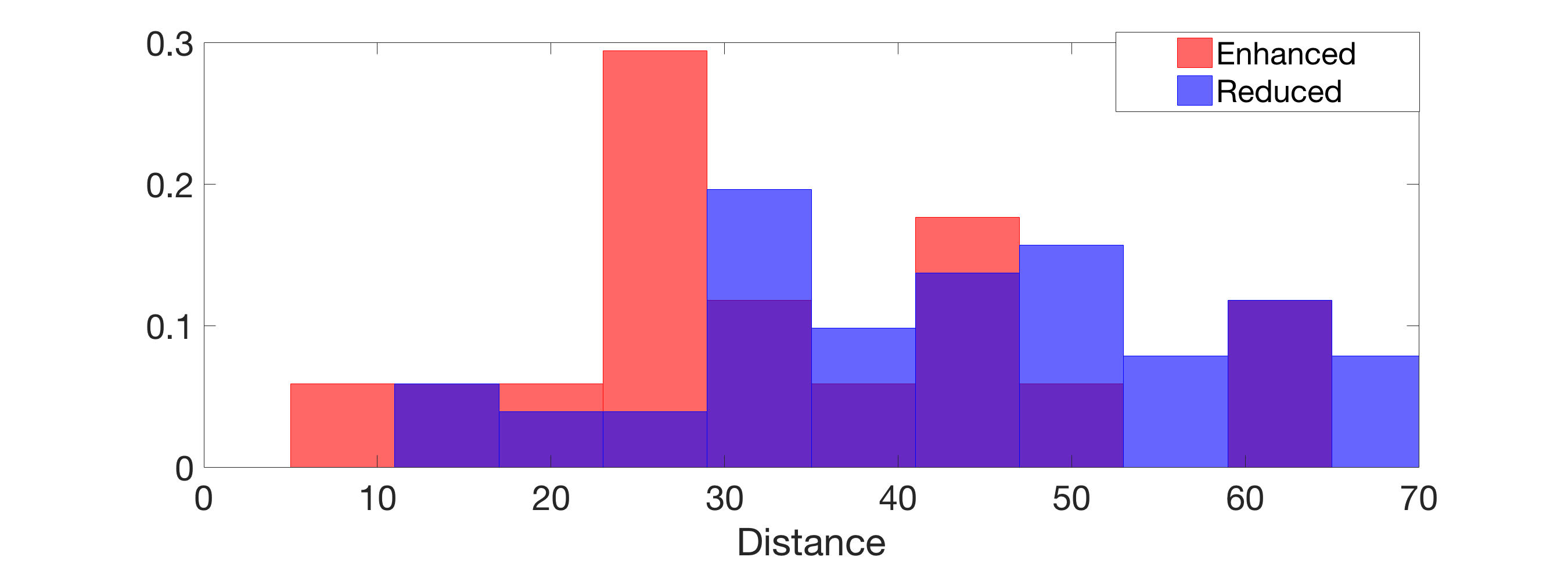}
		%\caption{Histogram of distance between ROIs for enhanced and reduced influences.}
		\caption{Relative frequency histogram of distances between ROIs for enhanced and reduced influences.}
		\label{Fig: ROI_distance}
	\end{figure}

	We compared the distance between paired ROIs with reduced and enhanced influences. Figure \ref{Fig: ROI_distance} gives a relative frequency histogram of distances between ROIs, measured by the Euclidean distance between the ROI coordinates, for enhanced and reduced influences. %shows the histogram of ROI distances for reduced influences and enhanced influences. Since the number of reduced and enhanced influences are different, we compared the frequency of observations. 
	We found that long rang influences are more likely to be reduced, while local influences are more likely to be enhanced. 
	
	%Figure 19 gives a relative frequency histogram of distances between ROIs, measured by the Euclidean distance between the ROI coordinates, for enhanced and reduced influences, demonstrating this trend.

	\subsection{Graph Properties}
	Furthermore, we investigated topological and geometrical properties of the estimated causal graphs from the three datasets. Particularly, we analyzed the following graph properties:
	\begin{itemize}
		\item Small-worldness. Small-world networks are characterized by dense local clustering of connections between neighboring nodes yet a short path length due to the existence of relatively few long-range connections. The small-world topology can support both segregated/specialized and distributed/integrated information processing \cite{Small_world1}. %Moreover, small-world networks are economical, tending to minimize wiring costs while supporting high dynamical complexity.
		Network small-worldness has been quantified by a small-world coefficient $s$, calculated by comparing clustering and path length of a given network to an equivalent random network with the same degree on average \cite{Small_world2, Small_world3}:
		\begin{equation*}
		s = \frac{C}{C_r} \cdot \frac{L_r}{L}
		\end{equation*}
		with $C$ and $C_r$ being the clustering coefficient of the target graph and the random graph, respectively, and $L$ and $L_r$ being the average shortest path length of the two graphs. In our experiments, we calculated $C$ and $L$ for directed graphs, since the causal network $B$ is directed, with the procedure in \cite{cc_implementation}. Specifically, for the average shortest path $L$, we considered two cases: one accounts for incident edges and the other accounts for outgoing edges. Thus, we have two characteristics of small-worldness, named in-small-worldness and out-small-worldness.
		\item Local clustering coefficient. It is a measure of the degree to which nodes in a graph tend to cluster together. Different from the global clustering coefficient used in the estimation of small-worldness, we considered local clustering coefficient (CC) for each ROI \cite{cc_implementation}. We also considered different components of the clustering coefficients corresponding to cycles, middles, in triangles, and out triangles \cite{cc_implementation}. For example, for the general CC, all possible directed triangles formed by each node are considered, no matter what the directions of their edges are; for CC corresponding to cycles, the accounted triangles form a cycle.
		\item Centrality. Indicators of centrality identify the most important nodes within a graph. We considered different characteristics of centrality, including
		\begin{itemize}
			\item Incloseness: the average length of the shortest path from all others node to the target node in the graph. Thus the more central a node is, the closer it is to all other nodes.
			\item Outcloseness: the average length of the shortest path from the target node to all other nodes in the graph.
			%\item Betweenness: the number of times a node acts as a bridge along the shortest path between two other nodes.
			\item Indegree: the number of links incident to the target node.
			\item Outdegree: the number of links out from the target node.
		\end{itemize}
		
	\end{itemize}
	
	Since the causal adjacency matrix $B$ is a real-valued matrix with $B_{jk} \in \mathcal{R}$, to estimate graph properties, we first set a threshold on $B_{jk}$ and derive the underlying causal graph $G$ with $G_{jk} \in \{0,1\}$. We considered different thresholds, $\alpha \in \{0.01,0.02,0.03,0.04,0.05,0.06,0.07,0.08,0.09,0.10\}$; if $B_{jk}\geq\alpha$, then $G_{jk} = 1$, and otherwise, $G_{jk} = 0$. We estimated both in- and out-small-worldness for each graph. Then a one-tailed two-sample t-test was performed to examine whether the small-worldness in the group of ASD is significantly smaller than that of TCs. Table \ref{Table: small-world} shows the $p$ values of in- and out-small-worldness on the three datasets with different thresholds. For the UM dataset, $p$ values are smaller than 0.05 (significance level) at all thresholds; for the USM dataset, $p$ values are smaller than 0.05 at threshold $ 0.01$  and $0.02$. The significantly smaller small-worldness indiactes that for individuals with ASD, their segregated/specialized and distributed/integrated information processing may be affected.% (more biological analysis)
	
	\begin{table}[htp!]
		\centering
		\begin{tabular}{lllllllllll}
			\multicolumn{11}{c}{(a) In-small-worldness}                                                                                                                                                                                                                                                                     \\ \hline
			\multicolumn{1}{|l|}{}    & \multicolumn{10}{c|}{Threshold}                                                                                                                                                                                                                                                       \\ \hline
			\multicolumn{1}{|l|}{}    & \multicolumn{1}{c|}{0.01} & \multicolumn{1}{l|}{0.02} & \multicolumn{1}{l|}{0.03} & \multicolumn{1}{l|}{0.04} & \multicolumn{1}{l|}{0.05} & \multicolumn{1}{l|}{0.06} & \multicolumn{1}{l|}{0.07} & \multicolumn{1}{l|}{0.08} & \multicolumn{1}{l|}{0.09} & \multicolumn{1}{l|}{0.10} \\ \hline
			\multicolumn{1}{|l|}{NYU} & \multicolumn{1}{l|}{0.82} & \multicolumn{1}{l|}{0.86} & \multicolumn{1}{l|}{0.71} & \multicolumn{1}{l|}{0.33} & \multicolumn{1}{l|}{0.25} & \multicolumn{1}{l|}{0.83} & \multicolumn{1}{l|}{0.86} & \multicolumn{1}{l|}{0.91} & \multicolumn{1}{l|}{0.68} & \multicolumn{1}{l|}{0.61} \\ \hline
			\multicolumn{1}{|l|}{UM}  & \multicolumn{1}{l|}{0.03} & \multicolumn{1}{l|}{0.00} & \multicolumn{1}{l|}{0.01} & \multicolumn{1}{l|}{0.01} & \multicolumn{1}{l|}{0.00} & \multicolumn{1}{l|}{0.01} & \multicolumn{1}{l|}{0.00} & \multicolumn{1}{l|}{0.00} & \multicolumn{1}{l|}{0.00} & \multicolumn{1}{l|}{0.00} \\ \hline
			\multicolumn{1}{|l|}{USM} & \multicolumn{1}{l|}{0.04} & \multicolumn{1}{l|}{0.02} & \multicolumn{1}{l|}{0.27} & \multicolumn{1}{l|}{0.18} & \multicolumn{1}{l|}{0.16} & \multicolumn{1}{l|}{0.25} & \multicolumn{1}{l|}{0.09} & \multicolumn{1}{l|}{0.11} & \multicolumn{1}{l|}{0.12} & \multicolumn{1}{l|}{0.16} \\ \hline
			&                           &                           &                           &                           &                           &                           &                           &                           &                           &                           \\
			\multicolumn{11}{c}{(b) Out-small-worldness}                                                                                                                                                                                                                                                                    \\ \hline
			\multicolumn{1}{|l|}{}    & \multicolumn{10}{c|}{Threshold}                                                                                                                                                                                                                                                       \\ \hline
			\multicolumn{1}{|l|}{}    & \multicolumn{1}{l|}{0.01} & \multicolumn{1}{l|}{0.02} & \multicolumn{1}{l|}{0.03} & \multicolumn{1}{l|}{0.04} & \multicolumn{1}{l|}{0.05} & \multicolumn{1}{l|}{0.06} & \multicolumn{1}{l|}{0.07} & \multicolumn{1}{l|}{0.08} & \multicolumn{1}{l|}{0.09} & \multicolumn{1}{l|}{0.10} \\ \hline
			\multicolumn{1}{|l|}{NYU} & \multicolumn{1}{l|}{0.79} & \multicolumn{1}{l|}{0.85} & \multicolumn{1}{l|}{0.67} & \multicolumn{1}{l|}{0.32} & \multicolumn{1}{l|}{0.26} & \multicolumn{1}{l|}{0.83} & \multicolumn{1}{l|}{0.84} & \multicolumn{1}{l|}{0.91} & \multicolumn{1}{l|}{0.68} & \multicolumn{1}{l|}{0.60} \\ \hline
			\multicolumn{1}{|l|}{UM}  & \multicolumn{1}{l|}{0.00} & \multicolumn{1}{l|}{0.01} & \multicolumn{1}{l|}{0.01} & \multicolumn{1}{l|}{0.00} & \multicolumn{1}{l|}{0.00} & \multicolumn{1}{l|}{0.01} & \multicolumn{1}{l|}{0.00} & \multicolumn{1}{l|}{0.00} & \multicolumn{1}{l|}{0.00} & \multicolumn{1}{l|}{0.00} \\ \hline
			\multicolumn{1}{|l|}{USM} & \multicolumn{1}{l|}{0.04} & \multicolumn{1}{l|}{0.02} & \multicolumn{1}{l|}{0.28} & \multicolumn{1}{l|}{0.18} & \multicolumn{1}{l|}{0.15} & \multicolumn{1}{l|}{0.23} & \multicolumn{1}{l|}{0.07} & \multicolumn{1}{l|}{0.08} & \multicolumn{1}{l|}{0.11} & \multicolumn{1}{l|}{0.18} \\ \hline
		\end{tabular}
		\caption{$P$ values of (a) in-small-worldness and (b) out-small-worldness on the three datasets.}
		\label{Table: small-world}
	\end{table}
	
	For characteristics of local clustering coefficients and centrality, they were estimated per ROI per subject. For example, let us denote the local clustering coefficient for the $i$th subject and $j$th ROI as $cc_j^{(i)}$. We first performed one-tailed two-sample t test on $\{cc_j^{(i)}\}_{i=1}^n$ for each ROI. To deal with multiple-testing problem, we did false discovery rate correction on the clustering coefficients of 116 ROIs at significance level 0.05. Table \ref{Table: Num_cc} shows the number of ROIs which are significantly increased/reduced in the group of ASD. We can see that on the UM dataset, there is a large number of reduction on all characteristics of local clustering coefficients, while there are also increases although fewer than reductions. For the NYU and USM datasets, there are only a few significant ROIs. %For UM dataset, most reductions are concentrated in......, while increases in........
	
	Table \ref{Table: Num_centrality} shows the number of ROIs with increased/reduced centrality in the group of ASD. Different characteristics of centrality are considered, including incloseness, outcloseness, indegree, and outdegree. For incloseness and outcloseness, on the UM dataset a great number of ROIs are reduced at most thresholds, which matches with the results from small-worldness, since closeness characterizes the shortest path length. On the USM dataset, the outclossness of a large number of ROIs are reduced at threshold $\alpha = 0.07, 0.08, 0.09, 0.10$. There are only a few significant ROIs on the NYU dataset, for both incloseness and outcloseness. For indegree, on all three datasets, only a few ROIs are reduced or increased. For outdegree, the number of ROIs which show reductions and increases are even at some particular threshold on all the three datasets.

	\begin{table}[htp!]
		\centering
		\begin{tabular}{lllllllllll}
			\multicolumn{11}{c}{(a) General clustering coefficient}                                                                                                                                                                                                                                                                           \\ \hline
			\multicolumn{1}{|l|}{}    & \multicolumn{10}{c|}{Threshold}                                                                                                                                                                                                                                                               \\ \hline
			\multicolumn{1}{|l|}{}    & \multicolumn{1}{l|}{0.01} & \multicolumn{1}{l|}{0.02} & \multicolumn{1}{l|}{0.03}   & \multicolumn{1}{l|}{0.04}   & \multicolumn{1}{l|}{0.05}  & \multicolumn{1}{l|}{0.06}  & \multicolumn{1}{l|}{0.07} & \multicolumn{1}{l|}{0.08} & \multicolumn{1}{l|}{0.09}  & \multicolumn{1}{l|}{0.10}  \\ \hline
			\multicolumn{1}{|l|}{NYU} & \multicolumn{1}{l|}{0/0}  & \multicolumn{1}{l|}{0/0}  & \multicolumn{1}{l|}{0/0}    & \multicolumn{1}{l|}{0/0}    & \multicolumn{1}{l|}{0/0}   & \multicolumn{1}{l|}{0/0}   & \multicolumn{1}{l|}{0/0}  & \multicolumn{1}{l|}{0/0}  & \multicolumn{1}{l|}{0/0}   & \multicolumn{1}{l|}{0/0}   \\ \hline
			\multicolumn{1}{|l|}{UM}  & \multicolumn{1}{l|}{0/4}  & \multicolumn{1}{l|}{0/9}  & \multicolumn{1}{l|}{0/18}   & \multicolumn{1}{l|}{0/1}    & \multicolumn{1}{l|}{0/2}   & \multicolumn{1}{l|}{44/72} & \multicolumn{1}{l|}{0/0}  & \multicolumn{1}{l|}{0/0}  & \multicolumn{1}{l|}{2/14}  & \multicolumn{1}{l|}{0/8}   \\ \hline
			\multicolumn{1}{|l|}{USM} & \multicolumn{1}{l|}{0/2}  & \multicolumn{1}{l|}{0/1}  & \multicolumn{1}{l|}{1/0}    & \multicolumn{1}{l|}{0/0}    & \multicolumn{1}{l|}{0/0}   & \multicolumn{1}{l|}{0/0}   & \multicolumn{1}{l|}{0/0}  & \multicolumn{1}{l|}{0/0}  & \multicolumn{1}{l|}{0/0}   & \multicolumn{1}{l|}{0/0}   \\ \hline
			&                           &                           &                             &                             &                            &                            &                           &                           &                            &                            \\
			\multicolumn{11}{c}{(b) Clustering coefficient (cycle)}                                                                                                                                                                                                                                                                     \\ \hline
			\multicolumn{1}{|l|}{}    & \multicolumn{10}{c|}{Threshold}                                                                                                                                                                                                                                                               \\ \hline
			\multicolumn{1}{|l|}{}    & \multicolumn{1}{l|}{0.01} & \multicolumn{1}{l|}{0.02} & \multicolumn{1}{l|}{0.03}   & \multicolumn{1}{l|}{0.04}   & \multicolumn{1}{l|}{0.05}  & \multicolumn{1}{l|}{0.06}  & \multicolumn{1}{l|}{0.07} & \multicolumn{1}{l|}{0.08} & \multicolumn{1}{l|}{0.09}  & \multicolumn{1}{l|}{0.10}  \\ \hline
			\multicolumn{1}{|l|}{NYU} & \multicolumn{1}{l|}{0/0}  & \multicolumn{1}{l|}{0/0}  & \multicolumn{1}{l|}{3/1}    & \multicolumn{1}{l|}{5/2}    & \multicolumn{1}{l|}{0/0}   & \multicolumn{1}{l|}{0/0}   & \multicolumn{1}{l|}{0/0}  & \multicolumn{1}{l|}{0/0}  & \multicolumn{1}{l|}{0/0}   & \multicolumn{1}{l|}{0/0}   \\ \hline
			\multicolumn{1}{|l|}{UM}  & \multicolumn{1}{l|}{0/2}  & \multicolumn{1}{l|}{4/73} & \multicolumn{1}{l|}{14/102} & \multicolumn{1}{l|}{12/104} & \multicolumn{1}{l|}{1/41}  & \multicolumn{1}{l|}{0/22}  & \multicolumn{1}{l|}{1/18} & \multicolumn{1}{l|}{0/3}  & \multicolumn{1}{l|}{0/13}  & \multicolumn{1}{l|}{1/17}  \\ \hline
			\multicolumn{1}{|l|}{USM} & \multicolumn{1}{l|}{0/2}  & \multicolumn{1}{l|}{0/0}  & \multicolumn{1}{l|}{0/1}    & \multicolumn{1}{l|}{0/2}    & \multicolumn{1}{l|}{0/0}   & \multicolumn{1}{l|}{0/0}   & \multicolumn{1}{l|}{0/0}  & \multicolumn{1}{l|}{0/1}  & \multicolumn{1}{l|}{0/2}   & \multicolumn{1}{l|}{0/0}   \\ \hline
			&                           &                           &                             &                             &                            &                            &                           &                           &                            &                            \\
			\multicolumn{11}{c}{(c) Clustering coefficient (middle)}                                                                                                                                                                                                                                                                    \\ \hline
			\multicolumn{1}{|l|}{}    & \multicolumn{10}{c|}{Threshold}                                                                                                                                                                                                                                                               \\ \hline
			\multicolumn{1}{|l|}{}    & \multicolumn{1}{l|}{0.01} & \multicolumn{1}{l|}{0.02} & \multicolumn{1}{l|}{0.03}   & \multicolumn{1}{l|}{0.04}   & \multicolumn{1}{l|}{0.05}  & \multicolumn{1}{l|}{0.06}  & \multicolumn{1}{l|}{0.07} & \multicolumn{1}{l|}{0.08} & \multicolumn{1}{l|}{0.09}  & \multicolumn{1}{l|}{0.10}  \\ \hline
			\multicolumn{1}{|l|}{NYU} & \multicolumn{1}{l|}{0/0}  & \multicolumn{1}{l|}{1/1}  & \multicolumn{1}{l|}{0/0}    & \multicolumn{1}{l|}{0/0}    & \multicolumn{1}{l|}{1/1}   & \multicolumn{1}{l|}{1/0}   & \multicolumn{1}{l|}{0/0}  & \multicolumn{1}{l|}{0/0}  & \multicolumn{1}{l|}{0/0}   & \multicolumn{1}{l|}{61/52} \\ \hline
			\multicolumn{1}{|l|}{UM}  & \multicolumn{1}{l|}{0/2}  & \multicolumn{1}{l|}{0/2}  & \multicolumn{1}{l|}{0/10}   & \multicolumn{1}{l|}{1/32}   & \multicolumn{1}{l|}{19/97} & \multicolumn{1}{l|}{0/11}  & \multicolumn{1}{l|}{7/48} & \multicolumn{1}{l|}{0/4}  & \multicolumn{1}{l|}{27/89} & \multicolumn{1}{l|}{29/87} \\ \hline
			\multicolumn{1}{|l|}{USM} & \multicolumn{1}{l|}{0/1}  & \multicolumn{1}{l|}{0/2}  & \multicolumn{1}{l|}{0/1}    & \multicolumn{1}{l|}{2/3}    & \multicolumn{1}{l|}{0/1}   & \multicolumn{1}{l|}{1/0}   & \multicolumn{1}{l|}{1/3}  & \multicolumn{1}{l|}{0/1}  & \multicolumn{1}{l|}{0/0}   & \multicolumn{1}{l|}{0/0}   \\ \hline
			&                           &                           &                             &                             &                            &                            &                           &                           &                            &                            \\
			\multicolumn{11}{c}{(d) Clustering coefficient (in triangles)}                                                                                                                                                                                                                                                              \\ \hline
			\multicolumn{1}{|l|}{}    & \multicolumn{10}{c|}{Threshold}                                                                                                                                                                                                                                                               \\ \hline
			\multicolumn{1}{|l|}{}    & \multicolumn{1}{l|}{0.01} & \multicolumn{1}{l|}{0.02} & \multicolumn{1}{l|}{0.03}   & \multicolumn{1}{l|}{0.04}   & \multicolumn{1}{l|}{0.05}  & \multicolumn{1}{l|}{0.06}  & \multicolumn{1}{l|}{0.07} & \multicolumn{1}{l|}{0.08} & \multicolumn{1}{l|}{0.09}  & \multicolumn{1}{l|}{0.10}  \\ \hline
			\multicolumn{1}{|l|}{NYU} & \multicolumn{1}{l|}{0/0}  & \multicolumn{1}{l|}{1/0}  & \multicolumn{1}{l|}{4/1}    & \multicolumn{1}{l|}{1/0}    & \multicolumn{1}{l|}{1/1}   & \multicolumn{1}{l|}{0/1}   & \multicolumn{1}{l|}{0/1}  & \multicolumn{1}{l|}{0/0}  & \multicolumn{1}{l|}{0/0}   & \multicolumn{1}{l|}{65/44} \\ \hline
			\multicolumn{1}{|l|}{UM}  & \multicolumn{1}{l|}{0/0}  & \multicolumn{1}{l|}{0/1}  & \multicolumn{1}{l|}{1/7}    & \multicolumn{1}{l|}{20/96}  & \multicolumn{1}{l|}{1/4}   & \multicolumn{1}{l|}{29/87} & \multicolumn{1}{l|}{0/0}  & \multicolumn{1}{l|}{0/2}  & \multicolumn{1}{l|}{0/5}   & \multicolumn{1}{l|}{0/1}   \\ \hline
			\multicolumn{1}{|l|}{USM} & \multicolumn{1}{l|}{0/2}  & \multicolumn{1}{l|}{0/2}  & \multicolumn{1}{l|}{0/1}    & \multicolumn{1}{l|}{0/1}    & \multicolumn{1}{l|}{0/0}   & \multicolumn{1}{l|}{0/0}   & \multicolumn{1}{l|}{0/0}  & \multicolumn{1}{l|}{0/0}  & \multicolumn{1}{l|}{0/0}   & \multicolumn{1}{l|}{0/0}   \\ \hline
			&                           &                           &                             &                             &                            &                            &                           &                           &                            &                            \\
			\multicolumn{11}{c}{(e) Clustering coefficient (out triangles)}                                                                                                                                                                                                                                                             \\ \hline
			\multicolumn{1}{|l|}{}    & \multicolumn{10}{c|}{Threshold}                                                                                                                                                                                                                                                               \\ \hline
			\multicolumn{1}{|l|}{}    & \multicolumn{1}{l|}{0.01} & \multicolumn{1}{l|}{0.02} & \multicolumn{1}{l|}{0.03}   & \multicolumn{1}{l|}{0.04}   & \multicolumn{1}{l|}{0.05}  & \multicolumn{1}{l|}{0.06}  & \multicolumn{1}{l|}{0.07} & \multicolumn{1}{l|}{0.08} & \multicolumn{1}{l|}{0.09}  & \multicolumn{1}{l|}{0.10}  \\ \hline
			\multicolumn{1}{|l|}{NYU} & \multicolumn{1}{l|}{0/0}  & \multicolumn{1}{l|}{2/1}  & \multicolumn{1}{l|}{5/1}    & \multicolumn{1}{l|}{1/0}    & \multicolumn{1}{l|}{0/0}   & \multicolumn{1}{l|}{0/0}   & \multicolumn{1}{l|}{0/0}  & \multicolumn{1}{l|}{0/0}  & \multicolumn{1}{l|}{1/0}   & \multicolumn{1}{l|}{0/0}   \\ \hline
			\multicolumn{1}{|l|}{UM}  & \multicolumn{1}{l|}{0/0}  & \multicolumn{1}{l|}{1/6}  & \multicolumn{1}{l|}{0/2}    & \multicolumn{1}{l|}{0/39}   & \multicolumn{1}{l|}{19/97} & \multicolumn{1}{l|}{0/23}  & \multicolumn{1}{l|}{1/3}  & \multicolumn{1}{l|}{0/0}  & \multicolumn{1}{l|}{0/3}   & \multicolumn{1}{l|}{0/0}   \\ \hline
			\multicolumn{1}{|l|}{USM} & \multicolumn{1}{l|}{0/0}  & \multicolumn{1}{l|}{0/0}  & \multicolumn{1}{l|}{0/0}    & \multicolumn{1}{l|}{0/1}    & \multicolumn{1}{l|}{0/1}   & \multicolumn{1}{l|}{1/1}   & \multicolumn{1}{l|}{1/1}  & \multicolumn{1}{l|}{0/0}  & \multicolumn{1}{l|}{0/0}   & \multicolumn{1}{l|}{47/69} \\ \hline
		\end{tabular}
		\caption{Number of ROIs with significantly increased/reduced characteristics of local clustering coefficients, including (a) general clustering coefficient and (b) clustering coefficient corresponding to cycles, (c) middles, (d) in triangles, and (e) out triangles.}
		\label{Table: Num_cc}
	\end{table}

	\begin{table}[htp!]
		\centering
		\begin{tabular}{lllllllllll}
			\multicolumn{11}{c}{(a) incloseness}                                                                                                                                                                                                                                                                           \\ \hline
			\multicolumn{1}{|l|}{}    & \multicolumn{10}{c|}{Threshold}                                                                                                                                                                                                                                                               \\ \hline
			\multicolumn{1}{|l|}{}    & \multicolumn{1}{l|}{0.01} & \multicolumn{1}{l|}{0.02} & \multicolumn{1}{l|}{0.03}   & \multicolumn{1}{l|}{0.04}   & \multicolumn{1}{l|}{0.05}  & \multicolumn{1}{l|}{0.06}  & \multicolumn{1}{l|}{0.07} & \multicolumn{1}{l|}{0.08} & \multicolumn{1}{l|}{0.09}  & \multicolumn{1}{l|}{0.10}  \\ \hline
			\multicolumn{1}{|l|}{NYU} & \multicolumn{1}{l|}{0/0}  & \multicolumn{1}{l|}{0/0}  & \multicolumn{1}{l|}{2/1}    & \multicolumn{1}{l|}{2/0}    & \multicolumn{1}{l|}{0/2}   & \multicolumn{1}{l|}{2/2}   & \multicolumn{1}{l|}{0/0}  & \multicolumn{1}{l|}{1/1}  & \multicolumn{1}{l|}{0/0}   & \multicolumn{1}{l|}{0/1}   \\ \hline
			\multicolumn{1}{|l|}{UM}  & \multicolumn{1}{l|}{0/0}  & \multicolumn{1}{l|}{0/0}  & \multicolumn{1}{l|}{0/9}   & \multicolumn{1}{l|}{0/37}    & \multicolumn{1}{l|}{14/102}   & \multicolumn{1}{l|}{1/75} & \multicolumn{1}{l|}{7/109}  & \multicolumn{1}{l|}{0/52}  & \multicolumn{1}{l|}{0/65}  & \multicolumn{1}{l|}{0/79}   \\ \hline
			\multicolumn{1}{|l|}{USM} & \multicolumn{1}{l|}{0/1}  & \multicolumn{1}{l|}{0/1}  & \multicolumn{1}{l|}{0/0}    & \multicolumn{1}{l|}{0/0}    & \multicolumn{1}{l|}{0/0}   & \multicolumn{1}{l|}{0/0}   & \multicolumn{1}{l|}{2/12}  & \multicolumn{1}{l|}{0/2}  & \multicolumn{1}{l|}{0/0}   & \multicolumn{1}{l|}{0/9}   \\ \hline
			&                           &                           &                             &                             &                            &                            &                           &                           &                            &                            \\
			\multicolumn{11}{c}{(b) outclossness}                                                                                                                                                                                                                                                                     \\ \hline
			\multicolumn{1}{|l|}{}    & \multicolumn{10}{c|}{Threshold}                                                                                                                                                                                                                                                               \\ \hline
			\multicolumn{1}{|l|}{}    & \multicolumn{1}{l|}{0.01} & \multicolumn{1}{l|}{0.02} & \multicolumn{1}{l|}{0.03}   & \multicolumn{1}{l|}{0.04}   & \multicolumn{1}{l|}{0.05}  & \multicolumn{1}{l|}{0.06}  & \multicolumn{1}{l|}{0.07} & \multicolumn{1}{l|}{0.08} & \multicolumn{1}{l|}{0.09}  & \multicolumn{1}{l|}{0.10}  \\ \hline
			\multicolumn{1}{|l|}{NYU} & \multicolumn{1}{l|}{0/0}  & \multicolumn{1}{l|}{0/0}  & \multicolumn{1}{l|}{0/0}    & \multicolumn{1}{l|}{1/1}    & \multicolumn{1}{l|}{0/0}   & \multicolumn{1}{l|}{0/0}   & \multicolumn{1}{l|}{0/0}  & \multicolumn{1}{l|}{0/0}  & \multicolumn{1}{l|}{0/0}   & \multicolumn{1}{l|}{0/0}   \\ \hline
			\multicolumn{1}{|l|}{UM}  & \multicolumn{1}{l|}{0/1}  & \multicolumn{1}{l|}{0/3} & \multicolumn{1}{l|}{0/0} & \multicolumn{1}{l|}{0/21} & 
			\multicolumn{1}{l|}{14/98}  & \multicolumn{1}{l|}{0/26}  & \multicolumn{1}{l|}{0/31} & \multicolumn{1}{l|}{2/44}  & \multicolumn{1}{l|}{0/56}  & \multicolumn{1}{l|}{6/110}  \\ \hline
			\multicolumn{1}{|l|}{USM} & \multicolumn{1}{l|}{0/0}  & \multicolumn{1}{l|}{0/0}  & \multicolumn{1}{l|}{0/2}    & \multicolumn{1}{l|}{0/0}    & \multicolumn{1}{l|}{0/0}   & \multicolumn{1}{l|}{0/1}   & \multicolumn{1}{l|}{1/11}  & \multicolumn{1}{l|}{14/102}  & \multicolumn{1}{l|}{0/14}   & \multicolumn{1}{l|}{0/10}   \\ \hline
			&                           &                           &                             &                             &                            &                            &                           &                           &                            &                            \\
			
			\multicolumn{11}{c}{(c) indegree}                                                                                                                                                                                                                                                                    \\ \hline
			\multicolumn{1}{|l|}{}    & \multicolumn{10}{c|}{Threshold}                                                                                                                                                                                                                                                               \\ \hline
			\multicolumn{1}{|l|}{}    & \multicolumn{1}{l|}{0.01} & \multicolumn{1}{l|}{0.02} & \multicolumn{1}{l|}{0.03}   & \multicolumn{1}{l|}{0.04}   & \multicolumn{1}{l|}{0.05}  & \multicolumn{1}{l|}{0.06}  & \multicolumn{1}{l|}{0.07} & \multicolumn{1}{l|}{0.08} & \multicolumn{1}{l|}{0.09}  & \multicolumn{1}{l|}{0.10}  \\ \hline
			\multicolumn{1}{|l|}{NYU} & \multicolumn{1}{l|}{0/0}  & \multicolumn{1}{l|}{0/0}  & \multicolumn{1}{l|}{2/0}    & \multicolumn{1}{l|}{1/0}    & \multicolumn{1}{l|}{0/1}   & \multicolumn{1}{l|}{2/3}   & \multicolumn{1}{l|}{10/3}  & \multicolumn{1}{l|}{0/1}  & \multicolumn{1}{l|}{0/0}   & \multicolumn{1}{l|}{0/1} \\ \hline
			\multicolumn{1}{|l|}{UM}  & \multicolumn{1}{l|}{0/0}  & \multicolumn{1}{l|}{0/0}  & \multicolumn{1}{l|}{0/1}   & \multicolumn{1}{l|}{0/1}   & \multicolumn{1}{l|}{1/1} & \multicolumn{1}{l|}{1/2}  & \multicolumn{1}{l|}{0/0} & \multicolumn{1}{l|}{0/1}  & \multicolumn{1}{l|}{0/3} & \multicolumn{1}{l|}{0/3} \\ \hline
			\multicolumn{1}{|l|}{USM} & \multicolumn{1}{l|}{0/1}  & \multicolumn{1}{l|}{0/1}  & \multicolumn{1}{l|}{0/0}    & \multicolumn{1}{l|}{0/0}    & \multicolumn{1}{l|}{0/0}   & \multicolumn{1}{l|}{1/0}   & \multicolumn{1}{l|}{0/0}  & \multicolumn{1}{l|}{0/0}  & \multicolumn{1}{l|}{0/2}   & \multicolumn{1}{l|}{0/4}   \\ \hline
			&                           &                           &                             &                             &                            &                            &                           &                           &                            &                            \\
			\multicolumn{11}{c}{(d) outdegree}                                                                                                                                                                                                                                                              \\ \hline
			\multicolumn{1}{|l|}{}    & \multicolumn{10}{c|}{Threshold}                                                                                                                                                                                                                                                               \\ \hline
			\multicolumn{1}{|l|}{}    & \multicolumn{1}{l|}{0.01} & \multicolumn{1}{l|}{0.02} & \multicolumn{1}{l|}{0.03}   & \multicolumn{1}{l|}{0.04}   & \multicolumn{1}{l|}{0.05}  & \multicolumn{1}{l|}{0.06}  & \multicolumn{1}{l|}{0.07} & \multicolumn{1}{l|}{0.08} & \multicolumn{1}{l|}{0.09}  & \multicolumn{1}{l|}{0.10}  \\ \hline
			\multicolumn{1}{|l|}{NYU} & \multicolumn{1}{l|}{0/0}  & \multicolumn{1}{l|}{0/0}  & \multicolumn{1}{l|}{0/0}    & \multicolumn{1}{l|}{0/0}    & \multicolumn{1}{l|}{0/0}   & \multicolumn{1}{l|}{0/0}   & \multicolumn{1}{l|}{4/0}  & \multicolumn{1}{l|}{59/57}  & \multicolumn{1}{l|}{0/0}   & \multicolumn{1}{l|}{0/1} \\ \hline
			\multicolumn{1}{|l|}{UM}  & \multicolumn{1}{l|}{0/1}  & \multicolumn{1}{l|}{0/0}  & \multicolumn{1}{l|}{0/0}    & \multicolumn{1}{l|}{0/0}  & \multicolumn{1}{l|}{0/2}   & \multicolumn{1}{l|}{0/2} & \multicolumn{1}{l|}{0/0}  & \multicolumn{1}{l|}{0/0}  & \multicolumn{1}{l|}{0/3}   & \multicolumn{1}{l|}{37/79}   \\ \hline
			\multicolumn{1}{|l|}{USM} & \multicolumn{1}{l|}{14/22}  & \multicolumn{1}{l|}{0/0}  & \multicolumn{1}{l|}{2/1}    & \multicolumn{1}{l|}{0/1}    & \multicolumn{1}{l|}{0/0}   & \multicolumn{1}{l|}{0/0}   & \multicolumn{1}{l|}{0/0}  & \multicolumn{1}{l|}{0/0}  & \multicolumn{1}{l|}{0/0}   & \multicolumn{1}{l|}{0/0}   \\ \hline
		\end{tabular}
		\caption{Number of ROIs with significantly increased/reduced characteristics of centrality, including (a) inclossness, (b) outcloseness, (c) indegree, and (d) outdegree.}
		\label{Table: Num_centrality}
	\end{table}
	
	\section{Discussion}
	The present study investigated atypical causal connections between brain regions for individuals with ASD. Particularly, we applied the two-step method to learn causal connectivities, which is statistically reliable in the case of high dimensionality and small sample size. The two-step method accounts for feedbacks and can identify the whole causal model. We examined our methods on three resting-state fMRI datasets. From experimental results, it shows that with causal connectivities, the out-of-sample diagnostic accuracy largely improves, compared with that using correlation and partial correlation as features for prediction. A closer examination shows that information flow (causal connection) starting from the superior front gyrus to other areas is largely reduced; the target areas are mostly concentrated in default mode network and the posterior part. Moreover, all enhanced information flows are from posterior to anterior or in local areas, while local reductions also exist. Overall, it shows that long-range connections have a larger proportion of reductions than local connections, while local connections have a larger proportion of increases than long-range connections. By examining the graph properties of brain causal structure, the group of ASD shows significantly reduced small-worldness compared to the control group on UM and USM datasets.
	
	%Unlike the present work which examines causal connections, there are a huge number of studies which documented functional connectivity with correlation or partial correlation. Correlation does not infer causation. With causality, it helps to pinpoint the targets more precisely, as well as considering the direction of information flow.
	
	The default mode network is known to be involved in the neurological basis for the self, thinking about others, remembering the past, and imagining the future \cite{DMN1}. Previous studies have shown that brain connections are disrupted, particularly reduced, in default mode network for individuals with ASD \cite{DMN2,DMN3, DMN4}. If the default mode network is altered, it may change the way one perceives events and their social and moral reasoning. Our findings are in concordance with these results. Particualrly, our study demonstrates the reduction from superior frontal gyrus to posterior cingulate gyrus, precuneus, angular gyrus, and parahippocampal gyrus. 
	
	The underconnectivity theory attributes the disorder to reduced anatomical connectivity and functional connectivity between the frontal cortex and more posterior areas of the brain \cite{Marcel_Just1, Marcel_Just2}. In our experiments, we pinpoint the reduced information flows  from superior frontal gyrus (medial orbital part) to posterior areas, such as ingual gyrus, temporal lobe, precuneus, and cerebellum. The superior frontal gyrus is shown to be involved in self-awareness, in coordination with the action of the sensory system \cite{Superior_frontal}. 
	
	Reduced long-range, relative to local, structural and functional connectivity of the brain has been demonstrated in autism by a large suite of studies \cite{Long_reduced1, Long_reduced2, Long_reduced3}. Local over-connectivity has also been found in several studies \cite{Local1, Local2}, but with conflicts in specific areas. It has been posited that integrated accounts of global under-connectivity and local over-connectivity may reinforce each other, resulting in failing to differentiate signal from noise \cite{Long_local1}.

\chapter{Conclusions and Future work}
In this paper, we studied the problem of ASD diagnosis by causal influence strength estimated from resting-state fMRI data. 

In Chapter 2, we provided a literature review on current diagnostic approaches of ASD, including behavior diagnosis, genetic testing, and neuroimaging-based diagnosis. Particularly, we focused on diagnosis from fMRI data. 

In Chapter 3, we focused on computational methods for causal discovery from fMRI data. We first gave a brief summary of recent developments in causal discovery. The limitations of current approaches for causal discovery from fMRI data thus motivate our proposed two-step method. The two-step method is able to recover the whole causal structure from measured signals reliably, including feedbacks among brain regions. It contains two steps. In the first step, it learns a superset of the underlying causal skeleton. The results from the first step are used as constraints of connectivities in the second step, which makes the estimation faster and more reliable, especially in the high-dimensional case. In the second step, it identifies the causal influence strength between ROIs for each subject. Particularly, the second step relies on constrained functional causal model, which represents the effect as a function of the direct causes together with an independent noise term.The resulting causal structure is uniquely identifiable under certain conditions. 

In Chapter 4, we applied the two-step method for causal connectivity estimation on ABIDE dataset, specifically, on datasets from three different sites which have the largest number of subjects. From the reported results, it shows that with causal connectivities, the out-of-sample diagnostic accuracy largely improves, compared with that using correlation and partial correlation as features for prediction. A closer examination shows that information flow starting from the superior front gyrus to other areas are largely reduced; the target areas are mostly concentrated in default mode network and the posterior part. Moreover, all enhanced information flows are from posterior to anterior or in local areas, while local reductions also exist. Overall, it shows that long-range influences have a larger proportion of reductions than local influences, while local influences have a larger proportion of increases than long-range influences. By examining the graph properties of brain causal structure, the group of ASD shows reduced small-worldness compared to the control group.

Future work for this line of research include the following aspects. First, for more reliable statistical analysis, larger homogeneous datasets are needed. Although the ABIDE dataset collects data from almost twenty sites, resulting in a large dataset, data distributions from different sites are heterogeneous. One of the reasons is that different machines and experimental protocols are used to collect the data. Because of the heterogeneity, one has to be careful when considering the datasets jointly. Second, ASD is a family of developmental disorders, containing several subtypes: autistic disorder, asperger syndrome, pervasive developmental disorder-not otherwise specified, and childhood disintegrative disorder. it is important to distinguish between different subtypes of ASD and identify their corresponding neurological mechanisms, which may provide a more accurate way for diagnosis and treatment. Finally, currently we use behavior assessment as ground truth to evaluate our proposed methods in fMRI data analysis. However, behavior assessment itself might not be accurate, and even the label information from behavior diagnosis may have noise. Thus, it is important to consider label noise in the trained model, or a better way is to develop reliable methods for clustering in high-dimensional case.

%\appendix
%\include{appendix}
    
\setcounter{chapter}{0}
\setcounter{table}{0}  
\chapter*{Appendix}
 \label{Appendix: ID}
\begin{table}[htp!]
		\caption{Subject IDs used in our experiments}
	\label{my-label}
	\centering
	\resizebox{0.75\columnwidth}{!}{
	\begin{tabular}{llllllllll}
		\multicolumn{10}{l}{Subject IDs in NYU dataset:}                               \\
		50952 & 50972 & 50991 & 51011 & 51029 & 51049 & 51066 & 51084 & 51102 & 51124 \\
		50954 & 50973 & 50992 & 51012 & 51030 & 51050 & 51067 & 51085 & 51103 & 51126 \\
		50955 & 50974 & 50993 & 51013 & 51032 & 51051 & 51068 & 51086 & 51104 & 51127 \\
		50956 & 50976 & 50994 & 51014 & 51033 & 51052 & 51069 & 51087 & 51105 & 51128 \\
		50957 & 50977 & 50995 & 51015 & 51034 & 51053 & 51070 & 51088 & 51106 & 51129 \\
		50958 & 50978 & 50996 & 51016 & 51035 & 51054 & 51072 & 51089 & 51107 & 51130 \\
		50959 & 50979 & 50997 & 51017 & 51036 & 51055 & 51073 & 51090 & 51109 & 51131 \\
		50960 & 50981 & 50999 & 51018 & 51038 & 51056 & 51074 & 51091 & 51110 &       \\
		50961 & 50982 & 51000 & 51019 & 51039 & 51057 & 51075 & 51093 & 51111 &       \\
		50962 & 50983 & 51001 & 51020 & 51040 & 51058 & 51076 & 51094 & 51112 &       \\
		50964 & 50984 & 51002 & 51021 & 51041 & 51059 & 51077 & 51095 & 51113 &       \\
		50965 & 50985 & 51003 & 51023 & 51042 & 51060 & 51078 & 51096 & 51114 &       \\
		50966 & 50986 & 51006 & 51024 & 51044 & 51061 & 51079 & 51097 & 51116 &       \\
		50967 & 50987 & 51007 & 51025 & 51045 & 51062 & 51080 & 51098 & 51117 &       \\
		50968 & 50988 & 51008 & 51026 & 51046 & 51063 & 51081 & 51099 & 51118 &       \\
		50969 & 50989 & 51009 & 51027 & 51047 & 51064 & 51082 & 51100 & 51122 &       \\
		50970 & 50990 & 51010 & 51028 & 51048 & 51065 & 51083 & 51101 & 51123 &       \\
		&       &       &       &       &       &       &       &       &       \\
		\multicolumn{10}{l}{Subject IDs in UM dataset:}                                \\
		50272 & 50285 & 50300 & 50318 & 50330 & 50340 & 50350 & 50358 & 50366 & 50375 \\
		50273 & 50291 & 50301 & 50319 & 50331 & 50341 & 50351 & 50359 & 50367 & 50376 \\
		50274 & 50292 & 50302 & 50320 & 50332 & 50342 & 50352 & 50360 & 50368 & 50377 \\
		50275 & 50293 & 50304 & 50321 & 50333 & 50343 & 50353 & 50361 & 50369 & 50379 \\
		50276 & 50294 & 50310 & 50324 & 50336 & 50344 & 50354 & 50362 & 50370 & 50380 \\
		50278 & 50295 & 50312 & 50325 & 50337 & 50345 & 50355 & 50363 & 50372 & 50381 \\
		50282 & 50297 & 50314 & 50327 & 50338 & 50348 & 50356 & 50364 & 50373 &       \\
		50284 & 50298 & 50315 & 50329 & 50339 & 50349 & 50357 & 50365 & 50374 &       \\
		&       &       &       &       &       &       &       &       &       \\
		\multicolumn{10}{l}{Subject IDs in USM dataset:}                               \\
		50477 & 50487 & 50496 & 50503 & 50515 & 50524 & 50532 & 50438 & 50445 & 50463 \\
		50480 & 50488 & 50497 & 50504 & 50516 & 50525 & 50432 & 50439 & 50446 & 50466 \\
		50481 & 50490 & 50498 & 50505 & 50518 & 50527 & 50433 & 50440 & 50447 & 50467 \\
		50482 & 50491 & 50499 & 50507 & 50519 & 50528 & 50434 & 50441 & 50448 & 50468 \\
		50483 & 50492 & 50500 & 50509 & 50520 & 50529 & 50435 & 50442 & 50449 & 50469 \\
		50485 & 50493 & 50501 & 50510 & 50521 & 50530 & 50436 & 50443 & 50453 & 50470 \\
		50486 & 50494 & 50502 & 50514 & 50523 & 50531 & 50437 & 50444 & 50456 &      
	\end{tabular}
}
\end{table}

\backmatter

%\renewcommand{\baselinestretch}{1.0}\normalsize

% By default \bibsection is \chapter*, but we really want this to show
% up in the table of contents and pdf bookmarks.

%\newcommand{\bibpreamble}{This text goes between the ``Bibliography''
%  header and the actual list of references}
\bibliography{reference_thesis}
\bibliographystyle{plainnat}

\end{document}